\documentclass[hidelinks,12pt]{article}
\usepackage[dvipsnames]{xcolor}
\usepackage{amsmath}
\usepackage{bbm}
\usepackage{bm}
\usepackage{amsmath, amsthm, amssymb}
\usepackage{setspace}
\usepackage{multirow}
\usepackage{graphicx}
\usepackage{natbib}
\usepackage{mathrsfs}
\usepackage{mdwlist}
\usepackage{rotating}
\usepackage{arydshln}
\usepackage{pdflscape}
\usepackage{mathtools}
\usepackage{enumitem}
\usepackage{todonotes}
\usepackage{pdflscape}
\usepackage{pbox}
\usepackage{fancyvrb}
\usepackage{rotating}
\usepackage{comment}
\usepackage{algorithmic}

\usepackage{mathtools,upgreek,bm}
\usepackage{times,epsfig,makeidx,amsfonts,dcolumn,multirow,graphicx,colortbl,url,enumitem}

\usepackage{booktabs}
\usepackage{pdfcomment}

\usepackage{caption}
\usepackage{subcaption}
\usepackage{lscape}

\hypersetup{
    colorlinks=false,
    pdfborder={0 0 0},
}

\theoremstyle{plain}

\usepackage{courier}
\normalfont
\usepackage[T1]{fontenc}

\makeindex

\newcommand{\captionfonts}{\footnotesize}
\makeatletter
\long\def\@makecaption#1#2{
  \vskip\abovecaptionskip
  \sbox\@tempboxa{{\captionfonts #1: #2}}
  \ifdim \wd\@tempboxa >\hsize
    {\captionfonts #1: #2\par}
  \else
    \hbox to\hsize{\hfil\box\@tempboxa\hfil}
  \fi
  \vskip\belowcaptionskip}
\makeatother

\newcommand{\diag}{\mathop{\mathrm{diag}}}

\newcommand{\beq}{\begin{equation}}
\newcommand{\eeq}{\end{equation}}

\newcommand{\bbeta}{\bm{\beta}}
\newcommand{\blambda}{\bm{\lambda}}

\newcommand{\defn}{\begin{quote}{\bf Definition. }}
\newcommand{\edefn}{\end{quote}}
\newcommand{\thm}{\begin{theorem}}
\newcommand{\ethm}{\end{theorem}}

\newcommand{\bmat}[1]{\left ( \begin{array}{#1}}
\newcommand{\emat}{\end{array}\right )}
\newcommand{\E}{\mathbb{E}}

\newcommand{\ts}{^{\sf T}}

\newcommand{\x}{ \textbf{x} }

\newcommand{\V}{ \textbf{V} }

\newcommand{\eb}{ \bm\epsilon }

\newcommand{\muz}{\bm\mu_{\textbf{M}}}

\theoremstyle{definition}

\theoremstyle{plain}
\newtheorem*{theorem}{Theorem}

\begin{document}

\title{Spline-Based Multi-State Models for Analyzing Disease Progression}

\author{Alessia Eletti \footnote{Department of Statistical Science, University College London, Gower Street, WC1E 6BT London, UK, \texttt{alessia.eletti.19@ucl.ac.uk}.} \and Giampiero Marra\footnote{Department of Statistical Science, University College London, Gower Street, WC1E 6BT London, UK, \texttt{giampiero.marra@ucl.ac.uk}.} \and Rosalba Radice \footnote{Faculty of Actuarial Science and Insurance, Bayes Business School, City, University of London, 106 Bunhill Row, EC1Y 8TZ London, UK, \texttt{rosalba.radice@city.ac.uk}.} }
\date{}

\maketitle

\vspace{-0.5cm}

\begin{abstract}
Motivated by disease progression-related studies, we propose an estimation method for fitting general non-homogeneous multi-state Markov models. The proposal can handle many types of multi-state processes, with several states and various combinations of observation schemes (e.g., intermittent, exactly observed, censored), and allows for the transition intensities to be flexibly modelled through additive (spline-based) predictors. The algorithm is based on a computationally efficient and stable penalized maximum likelihood estimation approach which exploits the information provided by the analytical Hessian matrix of the model log-likelihood. The proposed modeling framework is employed in case studies that aim at modeling the onset of cardiac allograft vasculopathy, and cognitive decline due to aging, where novel patterns are uncovered. To support applicability and reproducibility, all developed tools are implemented in the \texttt{R} package \texttt{flexmsm}.

\vspace{0.4cm}

\noindent \textbf{Keywords:} disease progression; information matrix; multi-state Markov model; penalized log-likelihood; penalized regression spline. 
\end{abstract}

\section{Introduction}\label{sec:introduction}

Continuous-time multi-state Markov models are the mainstream tools employed to model the progression of a disease through multiple stages, while accounting for the background information of each individual throughout the follow-up period \citep[e.g.,][]{gorfine2021marginalized, williams2020bayesian, yiu2017exploring}. Constant monitoring of a disease is not generally possible since it may be too expensive or not feasible due to the nature of the event of interest. In this case, the process is only observed at a fixed set of times and is thus said to be intermittently observed or interval-censored, a feature that makes the modeling problem harder. In applied studies, a non-homogeneous Markov process is often and rightly assumed, i.e. the risks of moving across states depend on the current state and on time. This is typically addressed by specifying parametric or more flexible (e.g., spline-based) models for the transition hazards \citep[e.g.,][]{cook2018multistate, gu2022maximum, joly2002penalized, machado2021penalised, sabathe2020regression, titman2011flexible, van2016multi}. 

We focus on two case studies, which motivated the development of the proposed methodology. Cardiac allograft vasculopathy (CAV) is a disease of the arterial walls which may arise in post-operative heart transplant recipients. Since death can be observed either with or without preceding CAV onset, the probability of death in each case is of interest. The focus is also on estimating the linear and potentially non-linear and interaction effects of relevant risk factors such as diagnosis that led to the transplant and donor age. The second study centres on cognitive decline due to aging, measured via a score-based system that leads to four separate degrees of severity, and the death event. This set up gives rise to five states. Here, interest lies in modeling linear or non-linear declining and improving trends, while accounting for covariate effects such as sex and schooling. As will be elaborated in the next paragraph, current software implementations did not allow us to fit the models required for the above modeling tasks.

The methodological literature on the subject is vast, however existing methods for fitting non-homogeneous multi-state Markov models are mainly based on the estimation approach developed by \cite{kalbfleisch1985analysis}, which relies on assuming piecewise-constant transition intensities, hence enabling the transition probabilities to be expressed as closed form functions of transition intensities, and on the use of the analytical score of the model log-likelihood for optimisation purposes. The advantage of the piecewise approximation is that, in principle, it makes it possible to develop a modeling framework that accounts for several states, forward and backward transitions, and any type of functional form for the transition intensities and covariate effects. However, only simple models can be fitted in practice, with the most commonly used software being the \texttt{R} package \texttt{msm} \citep{msm}. This implementation allows for several states, forward and backward transitions, but is limited in the specification for the transition intensities, since it focuses on either time-constant models or standard survival distributions (e.g., exponential and Gompertz). For the CAV study, only a model with log-linear effects of time and risk factors could be fitted. For the cognitive study, convergence could only be achieved for the trivial specification of a time-homogeneous process with no covariate effects. A bespoke \texttt{R} code for the well-known three-state Illness-Death Model (IDM) with a penalized smooth function of time and parametric covariate effects constrained to be equal across the transitions was provided by \cite{machado2021penalised}. When applied to the CAV study, this approach was found to be too restrictive for the estimation of covariate effects. Another implementation is given via the \texttt{R} package \texttt{nhm} \citep{nhm}. Here, the transition probabilities are obtained as numerical solutions of the differential equations that ties them to the transition intensities \citep{titman2011flexible}. This package is as general as \texttt{msm} but it additionally supports the use of an unpenalized smooth function of time. When applied to the CAV data, convergence could only be achieved for a model with log-linear effects. For the cognitive study, no model could be fitted.

Extensive experimentation with several custom-built routines based on various optimization schemes, exploiting up to first order analytical derivatives, showed that convergence can not be achieved when fitting the models required for the above studies. This finding extended to additional experiments based on a number of simulated datasets characterized by different numbers of states, backward and forward transitions, and specifications for the transition intensities and covariate effects. In all the experiments but those related to the simple models mentioned in the previous paragraph, it was consistently found that the algorithms are unable to propose sensible parameter values throughout the iterations, because of the absence of information on the local curvature of the transition probability matrix. Using numerical differentiation techniques produced poorly approximated information matrices, with the obvious consequences on convergence.

In this article, we propose an estimation method that exploits the information provided by the analytical Hessian matrix of the model log-likelihood. As opposed to the existing approaches, the advantage of the proposal lies in the flexibility offered for empirical modeling. It indeed allows for many types of multi-state processes, with several states, forward and backward transitions and various combinations of observation schemes (e.g., intermittent, exactly observed, censored), and for the transition intensities to be modelled through additive (spline-based) predictors. The use of the modeling framework is facilitated by the \texttt{R} package \texttt{flexmsm} \citep{flexmsm}, illustrated in the Supplementary Material.

In Section \ref{sec:math-setting}, we introduce multi-state Markov processes with transition intensities modelled through regression splines. The penalized log-likelihood is presented in Section \ref{sec:penal-lik}, while parameter estimation and how this is intertwined with the problem of computing the transition probabilities from the transition intensities are discussed in Section \ref{sec:par-estimation}. This section also presents the closed-form expressions for the transition probability matrix and its first and second derivatives, which are needed to compute the analytical likelihood, gradient and Hessian exploited in estimation. Section \ref{sec:inference} shows how inference is carried out, and Section \ref{sec:case-studies} describes and illustrates the two motivating case studies. Section \ref{sec:discussion} concludes the paper with some directions for future research. On-line Supplementary Materials \ref{suppl:log-lik-grad-Hess}, \ref{suppl:code} and \ref{suppl:algorithm} provide details on the model log-likelihood contributions and the use of the \texttt{R} package \texttt{flexmsm}. Supplementary Material \ref{suppl:sim-study} illustrates the empirical effectiveness of the proposal via two simulation studies. Supplementary Material \ref{suppl:symbols} contains a list of the mathematical symbols used and their meaning.

\section{Multi-state processes with spline-based transition intensities}\label{sec:math-setting}

Let $\{ Z(t), t > 0 \}$ be a continuous-time Markov process, $\mathcal{S} = \{ 1, 2, \dots, C\}$ its discrete state space, where $C$ is the total number of states, and $\mathcal{A} = \{ (r,r') \mid  r \neq r' \in \mathcal{S}, \exists \ r \rightarrow r' \}$ the set of transitions. The transition intensity function, i.e. the instantaneous rate of transition to a state $r'$ for an individual who is currently in another state $r$, is defined as follows
\begin{align*}
    q^{(rr')}(t) = \lim\limits_{h \downarrow 0} \dfrac{P(Z(t + h) = r' \mid Z(t) = r ) }{h}, \quad r \neq r',
\end{align*}
with $ q^{(rr')}(t) = 0$ if \textit{r} is an absorbing state and $ q^{(rr)}(t) = - \sum\limits_{r \neq r'}   q^{(rr')}(t)$. The matrix with $(r,r')$ element given by $q^{(rr')}(t)$ for every $r,r' \in \mathcal{S}$ is called transition intensity matrix or generator matrix and can be denoted with $\textbf{Q}(t)$. Similarly, the transition probability matrix associated with the time interval $(t,t')$ is defined as the matrix with $(r,r')$ element given by $p^{(rr')}(t,t') = P(Z(t') = r' \mid Z(t) = r)$ and can be denoted with $\textbf{P}(t, t')$. Here, we assume a time-dependent process as opposed to the rather restrictive time-homogeneous process (i.e., $\textbf{Q}(t) = \textbf{Q}$ $\forall t > 0$) often adopted in the literature for mathematical convenience.

The intensity for transition $r \rightarrow r'$, with $r \neq r'$, is generally represented using the proportional hazards specification, where the baseline intensity is specified using the exponential or Gompertz distribution \citep{van2016multi}. A more flexible representation for the transition intensity is
\begin{align}\label{ourin}
    q^{(rr')}(t_{\iota}) = \exp\left[ \eta^{(rr')}_{\iota}(t_{\iota},\textbf{x}_{\iota}; \bbeta^{(rr')}) \right],
\end{align}
 where $t_\iota$ and $\textbf{x}_\iota$ are the time and the vector of characteristics for observation $\iota$ respectively, $\bbeta^{(rr')}$ is the associated regression coefficient vector and $\eta^{(rr')}_{\iota}(t_{\iota},\textbf{x}_{\iota}; \bbeta^{(rr')}) \in \mathbb{R}$ is an additive predictor, discussed in detail in the following section, which includes a baseline smooth function of time and several types of covariate effects.

\subsection{Additive predictor}\label{sec:additive-predictor}

For simplicity, the dependence on covariates and parameters has been dropped when discussing the construction of $\eta^{(rr')}_{\iota}$. 

An additive predictor allows for various types of covariate effects and is defined as 
\begin{align}\label{linpred}
    \eta^{(rr')}_{\iota} = \beta_{0}^{(rr')} + \sum_{k=1}^{K^{(rr')}} s_{k}^{(rr')}(\tilde{\textbf{x}}_{k \iota}), \quad \iota = 1, \ldots, \check{n},
\end{align}
where $\check{n}$ is the sample size, $\beta_{0}^{(rr')} \in \mathbb{R}$ is an overall intercept, $\tilde{\textbf{x}}_{k \iota}$ denotes the $k^{th}$ sub-vector of the complete vector $\tilde{\textbf{x}}_{\iota} = (t_\iota, \x_\iota\ts)\ts$ and the $K^{(rr')}$ functions $s_{k}^{(rr')}(\tilde{\textbf{x}}_{k \iota})$ represent effects which are chosen according to the type of covariate(s) considered. For example, if we were interested in modeling a time-dependent effect of the covariate $age_\iota$, then $\tilde{\textbf{x}}_{\iota k}$ would be the vector $(age_\iota, t_\iota)\ts$ and $s_{k}^{(rr')}(age_\iota, t_\iota)$ the corresponding joint effect. 

Each $s_{k}^{(rr')}(\tilde{\textbf{x}}_{k \iota})$ can be represented as a linear combination of $J_{k}^{(rr')}$ known basis functions $\textbf{b}^{(rr')}_{k}(\tilde{\textbf{x}}_{k \iota}) = \left(b^{(rr')}_{k 1}(\tilde{\textbf{x}}_{k \iota}), \dots, b^{(rr')}_{k J_k^{(rr')}}(\tilde{\textbf{x}}_{k \iota})  \right)\ts$ and regression coefficients $\bbeta^{(rr')}_k = \left(\beta^{(rr')}_{k 1}, \ldots , \beta^{(rr')}_{k J_k^{(rr')}} \right)\ts \in \mathbb{R}^{J_k^{(rr')}}$, that is $s_{k}^{(rr')}(\tilde{\textbf{x}}_{k \iota}) = \textbf{b}^{(rr')}_{k}(\tilde{\textbf{x}}_{k \iota}) \ts \bbeta^{(rr')}_k $ \citep[e.g.,][]{Wood}.
The above formulation implies that the vector of evaluations $\left\{s^{(rr')}_{k}(\tilde{\textbf{x}}_{k 1}), \ldots, s^{(rr')}_{k}(\tilde{\textbf{x}}_{k \check{n}})\right\}\ts$ can be written as $\tilde{\textbf{X}}^{(rr')}_k \bbeta^{(rr')}_k$, where $\tilde{\textbf{X}}^{(rr')}_k$ is the design matrix whose $\iota^{th}$ row is given by $\textbf{b}^{(rr')}_{k}(\tilde{\textbf{x}}_{k \iota})\ts$ for $\iota = 1, \dots, \check{n}$. This allows the predictor in equation (\ref{linpred}) to be written as $\boldsymbol{\eta}^{(rr')} = \beta_{0}^{(rr')} \textbf{1}_{\check{n}} + \tilde{\textbf{X}}_{1}^{(rr')} \bbeta_{1}^{(rr')} + \ldots + \tilde{\textbf{X}}_{K^{(rr')}}^{(rr')} \bbeta_{K^{(rr')}}^{(rr')}$, where $\textbf{1}_{\check{n}}$ is an $\check{n}$-dimensional vector made up of ones. This can also be represented in a more compact way as $\boldsymbol{\eta}^{(rr')} = \tilde{\textbf{X}}^{(rr')} \bbeta^{(rr')}$, where $\tilde{\textbf{X}}^{(rr')} = (\mathbf{1}_n,\tilde{\textbf{X}}_{1}^{(rr')},\ldots,\tilde{\textbf{X}}_{K^{(rr')}}^{(rr')})$ and $\bbeta^{(rr')} = \left(\beta_{0}^{(rr') \sf T}, \bbeta_{1}^{(rr') \sf T}, \ldots, \bbeta_{K^{(rr')}}^{(rr') \sf T}\right)\ts$. Each $\bbeta_{k}^{(rr')}$ has an associated quadratic penalty $\lambda_{k}^{(rr')} \bbeta_{k}^{(rr') \sf T} \textbf{D}^{(rr')}_{k} \bbeta^{(rr')}_{k}$, used in fitting, whose role is to enforce specific properties on the $k^{th}$ function, such as smoothness. Matrix $\textbf{D}^{(rr')}_{k}$ only depends on the choice of the basis functions. Smoothing parameter $\lambda^{(rr')}_{k} \in [0,\infty)$ has the crucial role of controlling the trade-off between fit and smoothness and hence it determines the shape of the corresponding estimated smooth function. The overall penalty is defined as $\bbeta^{(rr') \sf T} \textbf{S}_{\boldsymbol\lambda^{(rr')}}^{(rr')} \bbeta^{(rr')}$, where $\textbf{S}^{(rr')}_{\boldsymbol\lambda^{(rr')}} = \diag(0, \lambda^{(rr')}_{1}\textbf{D}^{(rr')}_{1}, \ldots, \lambda^{(rr')}_{K^{(rr')}}\textbf{D}^{(rr')}_{K^{(rr')}})$ and $\boldsymbol\lambda^{(rr')} = ( \lambda^{(rr')}_{1}, \ldots, \lambda^{(rr')}_{K^{(rr')}})\ts$ is the transition-specific overall smoothing parameter vector. Note that smooth functions are subject to centering (identifiability) constraints which are imposed as described in \citet[][Section 5.4.1]{Wood}. Several definitions of basis functions and penalty terms are supported by \texttt{flexmsm}; these include thin plate, cubic and P-regression splines, tensor products, Markov random fields, random effects, and Gaussian process smooths \citep[][Chapter 5]{Wood}. 

An example of predictor specification is $\eta^{(rr')}_{\iota} = \beta_{0}^{(rr')} + s_1^{(rr')}(t_\iota) + \beta_2^{(rr')} sex_\iota$. Parametric effects usually, but not exclusively, relate to binary and categorical variables such as $sex_\iota$. The spline representation introduced above thus simplifies to $s_2^{(rr')}(sex_\iota) = \beta_2^{(rr')} sex_\iota$. No penalty is typically assigned to parametric effects, hence the associated penalty is $0$. However, there might be instances where some form of regularisation is required in which case a suitable penalization scheme can be employed \citep[e.g.,][Section 5.8]{Wood}. To explore a potentially nonlinear effect of $t_\iota$, $s_1^{(rr')}(t_\iota)$ is specified as $\textbf{b}_{1}^{(rr')}(t_\iota)\ts \bbeta_{1 }^{(rr')}$, where $\textbf{b}_{1}^{(rr')}(t_\iota)$ are cubic regression spline bases, for example. In this case, the penalty is defined as
\begin{align*}
    \boldsymbol{\beta}_{1}^{(rr')\ts} \textbf{D}^{(rr')}_1 \boldsymbol{\beta}_{1}^{(rr')} = \int_{u_1}^{u_{J_1^{(rr')}}} \left( \frac{\partial^2}{\partial t^2} s_1^{(rr')}(u) \right)^2  d u,
\end{align*}
where $u_1$ and $u_{J_1^{(rr')}}$ are the locations of the first and last knots. For a smooth term in one dimension, such as $s_1^{(rr')}(t_\iota)$, the specific choice of spline definition (e.g., thin plate, cubic) will not have an impact on the estimated curve. As for $J_1^{(rr')}$, or more generally $J_k^{(rr')}$, this is typically set to 10 since such value offers enough flexibility in most applications. However, analyzes using larger values can be attempted to assess the sensitivity of the results to $J_k^{(rr')}$. Regarding the selection of knots, these can be placed evenly throughout (or using the percentiles of) the values of the variable the smooth term refers to. For a thin-plate regression spline only $J_k^{(rr')}$ has to be chosen. See \citet[][Chapter 5]{Wood} for a thorough discussion. 

As mentioned previously, our framework poses no limits on the types of splines that can be employed for specifying the transition intensities. For instance, as illustrated in Section \ref{sec:CAV-study}, two-dimensional splines can be used to incorporate time-dependent effects. This would take the form of an interaction term involving, e.g., $age_\iota$ and the time variable through the smooth term $s^{(rr')}_k(age_\iota, t_\iota)$. Here, we have two penalties, one for each of the arguments of the smooth function. These are summed after being weighted by smoothing parameters, which serve the purpose of controlling the trade-off between fit and smoothness in each of the two directions, thus allowing for a great degree of flexibility \citep[][Section 5.6]{Wood}.

\section{Penalized log-likelihood}\label{sec:penal-lik}

Let $N$ be the number of statistical units, $n_i$ the number of times the $i^{th}$ unit is observed, $0 = t_{i0} < t_{i1} < \dots < t_{i n_i}$ the follow-up times, $z_{i0}, z_{i1}, \dots, z_{i n_i}$ the (possibly unobserved, i.e. censored) states occupied, and $\check{n} = \sum_{i = 1}^{N} (n_i - 1)$ the sample size. If $L_{i j} (\boldsymbol\theta)$ is the likelihood contribution for the $j^{th}$ observation of the $i^{th}$ unit and $\boldsymbol\theta = \{ \bbeta^{(rr')} \mid (r , r') \in \mathcal{A} \}$ the model parameter vector, then the log-likelihood is
\begin{align}\label{eq:general-lik}
    \ell(\boldsymbol\theta) = \sum\limits_{i = 1}^{N} \sum\limits_{j = 1}^{n_i} \log\left(L_{i j} (\boldsymbol\theta)\right),
\end{align}
where we have
\begin{align*}
    L_{ij}(\boldsymbol{\theta}) = 
    \begin{cases}
   p^{(z_{i j-1} z_{i j})}(t_{i j-1}, t_{i j}), \quad &\text{if $z_{i j}$ is an interval censored state} \vspace{0.1cm} \\ 
    \exp\bigg[ \int\limits_{t_{i j-1}}^{t_{ij}} q^{(z_{i j-1} z_{i j-1})}(u) du \bigg] q^{(z_{i j-1} z_{i j})}(t_{i j}), \quad &\text{if $z_{i j}$ is an exactly observed state} \vspace{0.1cm} \\ 
    \sum\limits_{ c \in \tilde{\mathcal{S}}} p^{(z_{i j-1} c)}(t_{i j-1}, t_{i j}), \quad &\text{ if $z_{i j}$ is a censored state} \vspace{0.1cm} \\
    \sum\limits_{\substack{c = 1 \\ c \neq z_{i j}}}^C p^{(z_{i j-1} c)}(t_{i j-1}, t_{i j}) q^{(c z_{i j})}(t_{i j}), \quad &\text{ if $z_{i j}$ is an exactly observed death state}
    \end{cases}.
\end{align*}
That is, the likelihood contribution for a given observation will depend on the nature of the states between which the transition occurred and the way in which it was observed. The first two contribution types refer, respectively, to the cases where a transition into a state is only known to have occurred in an interval $(t_{ij-1}, t_{ij})$ and when the transition time is known exactly. The third contribution accounts for the case where a transition is known to have occurred between $t_{i j-1}$ and $t_{ij-1}$, but the state occupied is only known to lie in a set $\tilde{\mathcal{S}} \subset \mathcal{S}$, for example, because the information was not registered or lost. The last contribution refers to a case that is common in observational studies of chronic diseases, i.e. when the time of death is known, but the state immediately before death is unknown. Supplementary Material \ref{suppl:log-lik-grad-Hess} provides further details on each contribution type, whereas Supplementary Material \ref{suppl:code} describes the use of the \texttt{R} package \texttt{flexmsm} in such a general context.

To calibrate the trade-off between parsimony and complexity, we augment the objective function (\ref{eq:general-lik}) with a quadratic penalty term. This results in the penalized log-likelihood
\begin{align}\label{Penloglik}
    \ell_p(\boldsymbol\theta) = \ell(\boldsymbol\theta) - \dfrac{1}{2} \boldsymbol\theta\ts \textbf{S}_{\boldsymbol\lambda} \boldsymbol\theta,
\end{align}
where $\textbf{S}_{\boldsymbol\lambda} = \diag\left( \{ \textbf{S}_{\boldsymbol\lambda^{(rr')}}^{(rr')} \mid (r , r') \in \mathcal{A}  \} \right)$ which is a block diagonal matrix where each block is given by the transition-specific penalty matrix $\textbf{S}^{(rr')}_{\boldsymbol\lambda^{(rr')}}$, and $\boldsymbol\lambda =  \{ \boldsymbol\lambda^{(rr')} \mid (r , r') \in \mathcal{A}  \}$ is the overall multiple smoothing parameter vector. Both $\textbf{S}^{(rr')}_ {\boldsymbol\lambda^{(rr')}}$ and $\boldsymbol\lambda^{(rr')}$ are defined for a generic transition $(r,r')$ in Section \ref{sec:additive-predictor}.

\section{Parameter estimation}\label{sec:par-estimation}

Model fitting is achieved by adapting to the current setting the stable and efficient approach proposed in \cite{Marra2019}, which combines a trust region algorithm with automatic multiple smoothing parameter selection. It is well known that the trust region method, when provided with first and second order analytical derivatives, significantly outperforms its line search counterparts and has optimal convergence properties \citep[Chapter 4,][]{Nocedal}. As for the smoothing parameters, we employ a theoretically founded framework that requires the availability of the analytical observed information matrix to achieve fast and stable estimation of $\boldsymbol\lambda$. More details are given in Supplementary Material \ref{suppl:algorithm}. Here, we focus on discussing the elements required to implement the above estimation approach for the proposed multi-state Markov modeling framework. 

From (\ref{eq:general-lik}), the $w^{th}$ element of the gradient vector $\textbf{g}(\boldsymbol\theta)$ and the $(w, w')$ element of the Hessian matrix $\textbf{H}(\boldsymbol\theta)$, for $w, w' = 1, \dots, W$ with $W = \sum_{(r,r') \in \mathcal{A}} \left( 1 + \sum_{k = 1}^{K^{(rr')}} J_k^{(rr')}\right)$, are defined as
\begin{align*}
    \dfrac{\partial }{\partial \theta_w} \ell(\boldsymbol\theta) & = \sum\limits_{i = 1}^{N} \sum\limits_{j = 1}^{n_i}  L_{i j}(\boldsymbol\theta)^{-1} \dfrac{\partial }{\partial \theta_w} L_{i j}(\boldsymbol\theta), \\
    \dfrac{\partial^2 }{\partial \theta_w \partial \theta_{w'}} \ell(\boldsymbol\theta) & = \sum\limits_{i = 1}^{N} \sum\limits_{j = 1}^{n_i} \bigg( L_{i j}(\boldsymbol\theta)^{-1} \dfrac{\partial^2 }{\partial \theta_w \partial \theta_{w'}} L_{i j}(\boldsymbol\theta) - L_{i j}(\boldsymbol\theta)^{-2} \dfrac{\partial }{\partial \theta_w} L_{i j}(\boldsymbol\theta) \dfrac{\partial }{\partial \theta_{w'}} L_{i j}(\boldsymbol\theta) \bigg),
\end{align*}
where $\dfrac{\partial L_{i j}(\boldsymbol\theta)}{\partial \theta_w}$ is given by
\begin{align*}
    \begin{cases}
        \dfrac{\partial }{\partial \theta_w} p^{(z_{i j-1} z_{i j})}(t_{i j-1}, t_{i j}), \quad &\text{if $z_{i j}$ is an interval censored state} \vspace{0.2cm} \\ 
         \exp\bigg( \int\limits_{t_{i j-1}}^{t_{ij}} q^{(z_{i j-1} z_{i j-1})}(u) du \bigg) \bigg[ \dfrac{\partial}{\partial \theta_w} q^{(z_{i j-1} z_{i j})}(t_{i j})  \quad &\text{if $z_{i j}$ is an exactly observed state} \\
         + q^{(z_{i j-1} z_{i j})}(t_{i j}) \int\limits_{t_{i j-1}}^{t_{ij}} \dfrac{\partial}{\partial \theta_w} q^{(z_{i j-1} z_{i j-1})}(u) du  \bigg], \vspace{0.2cm} \\
        \sum\limits_{c = 1}^C \dfrac{\partial }{\partial \theta_w} p^{(z_{i j-1} c)}(t_{i j-1}, t_{i j}), \quad &\text{if $z_{i j}$ is a censored state} \vspace{0.2cm} \\
        \sum\limits_{c = 1}^C \dfrac{\partial  }{\partial \theta_w} p^{(z_{i j-1} c)}(t_{i j-1}, t_{i j}) q^{(c z_{i j})}(t_{i j}) \quad &\text{if $z_{i j}$ is an exactly observed death state}  \\
        \hspace{0.8cm} + p^{(z_{i j-1} c)}(t_{i j-1}, t_{i j}) \dfrac{\partial }{\partial \theta_w}  q^{(c z_{i j})}(t_{i j}), 
    \end{cases},
\end{align*}
and $ \dfrac{\partial^2 L_{i j}(\boldsymbol\theta)}{\partial \theta_w \partial \theta_{w'}}$ is given by
\begin{align*}
    \begin{cases}
        \dfrac{\partial^2 }{\partial \theta_w \partial \theta_{w'}} p^{(z_{i j-1} z_{i j})}(t_{i j-1}, t_{i j}), \quad &\text{if $z_{i j}$ is an interval censored state} \vspace{0.2cm} \\ 
        \dfrac{\partial }{\partial \theta_w} L_{i j}(\boldsymbol\theta) \int\limits_{t_{i j-1}}^{t_{ij}} \dfrac{\partial}{\partial \theta_{w'}} q^{(z_{i j-1} z_{i j-1})}(u) du \quad &\text{if $z_{i j}$ is an exactly observed state}  \\
        + \exp\bigg( \int\limits_{t_{i j-1}}^{t_{ij}} q^{(z_{i j-1} z_{i j-1})}(u) du \bigg) \bigg[ \dfrac{\partial^2 q^{(z_{i j-1} z_{i j})}(t_{i j})}{\partial \theta_w \partial \theta_{w'}}  \\
         + \dfrac{\partial}{\partial \theta_{w'}} q^{(z_{i j-1} z_{i j})}(t_{i j}) \int\limits_{t_{i j-1}}^{t_{ij}} \dfrac{\partial}{\partial \theta_w} q^{(z_{i j-1} z_{i j-1})}(u) du \\
         + q^{(z_{i j-1} z_{i j})}(t_{i j}) \int\limits_{t_{i j-1}}^{t_{ij}} \dfrac{\partial^2 q^{(z_{i j-1} z_{i j-1})}(u)}{\partial \theta_w \partial \theta_{w'}}  du \bigg], \vspace{0.3cm} \\ 
        \sum\limits_{c = 1}^C \dfrac{\partial^2 }{\partial \theta_w \partial \theta_{w'}} p^{(z_{i j-1} c)}(t_{i j-1}, t_{i j}), \quad &\text{if $z_{i j}$ is a censored state} \vspace{0.2cm} \\
         \sum\limits_{c = 1}^C \dfrac{\partial^2  }{\partial \theta_w \partial \theta_{w'}} p^{(z_{i j-1} c)}(t_{i j-1}, t_{i j}) q^{(c z_{i j})}(t_{i j}) \quad &\text{if $z_{i j}$ is an exactly observed death state}   \\
         \hspace{0.8cm} + \dfrac{\partial  }{\partial \theta_w} p^{(z_{i j-1} c)}(t_{i j-1}, t_{i j}) \dfrac{\partial }{\partial \theta_{w'}}  q^{(c z_{i j})}(t_{i j}) \\
         \hspace{0.8cm} + \dfrac{\partial }{\partial \theta_{w'}} p^{(z_{i j-1} c)}(t_{i j-1}, t_{i j}) \dfrac{\partial }{\partial \theta_w}  q^{(c z_{i j})}(t_{i j}) \\
         \hspace{0.8cm} +  p^{(z_{i j-1} c)}(t_{i j-1}, t_{i j}) \dfrac{\partial^2 }{\partial \theta_w \partial \theta_{w'}}  q^{(c z_{i j})}(t_{i j}), 
    \end{cases}.
\end{align*}
The quantities needed for parameter estimation are the $C \times C$ dimensional matrices $\textbf{P}(t_{i j-1}, t_{i j})$, $\partial \textbf{P}(t_{i j-1}, t_{i j}) / \partial \theta_w$ and $\partial^2 \textbf{P}(t_{i j-1}, t_{i j}) / \partial \theta_w \partial \theta_{w'}$ for $w, w' = 1, \dots, W$. 

\subsection{Computation of the transition probability matrix and its first and second derivatives}\label{sec:trans-prob-computation}





Given the transition intensity matrix $\textbf{Q}(t)$, the transition probability matrix is the solution of the Kolmogorov forward differential equations $\partial \textbf{P}(t,t') / \partial t' = \textbf{P}(t,t') \textbf{Q}(t')$, which are not in general tractable. 
To tackle this, we employ the commonly adopted piecewise-constant approximation approach, with time grid coinciding with the observation times of the dataset at hand \citep{van2016multi}. Note that grids can be defined differently if required \citep[e.g.,][]{van2008multi}. For each individual $i = 1, \dots, N$, let the observed follow-up times $t_{i0} < t_{i1} < \dots < t_{i n_i}$ define the extremities of the intervals over which the transition intensities are assumed to be constant. The convention is to assume that the transition intensities remain constant on the value taken in the left extremity of each time interval. Then, for $t \in [t_{ij}, t_{i j+1})$, with $j = 0, 1, \dots, n_i-1$, making explicit the dependence on the model parameters, we have $\textbf{Q}(t; \boldsymbol\theta) = \textbf{Q}_j(\boldsymbol\theta)$ and
\begin{align}\label{eq:P-exp-Q}
    \textbf{P}(t_{ij}, t_{i j+1}) = \textbf{P}(t_{i j+1} - t_{i j}) = \exp[(t_{i j+1} - t_{i j}) \textbf{Q}_j(\boldsymbol\theta) ] = \sum\limits_{\zeta = 0}^{\infty} \dfrac{[(t_{i j+1} - t_{i j}) \textbf{Q}_j(\boldsymbol\theta)]^\zeta}{\zeta!}.
\end{align}
It follows that computing the transition probability matrix and its derivatives entails calculating a number of matrix exponentials and their derivatives. 
The eigendecomposition approach popularised by \cite{kalbfleisch1985analysis} here is appealing because it provides a closed-form solution for these power series. The availability of a closed-form expression is crucial since solving the power series for the transition probability matrix and its derivatives is a rather involved process, due to the matrix-multiplication being non-commutative. In particular, the authors provide analytical expressions for $\textbf{P}(t_{i j-1}, t_{i j})$ and $\partial \textbf{P}(t_{i j-1}, t_{i j}) / \partial \theta_w$, but not for $\partial^2 \textbf{P}(t_{i j-1}, t_{i j}) / \partial \theta_w \partial \theta_{w'}$, which is needed to derive the observed information matrix, and only when the eigenvalues of the transition intensity matrix are distinct. Much of the literature on interval-censored multi-state process followed this work and thus relies on up to first order information only.

A lesser known and so far unexploited result is that by \cite{kosorok1996analysis}, who provide a closed-form solution for $\partial^2 \textbf{P}(t_{i j}, t_{i j+1}) / \partial \theta_w \partial \theta_{w'}$. From this work it also emerged that the derived expressions do not require the eigenvalues of the transition intensity matrix to be distinct, which was not noted in \cite{kalbfleisch1985analysis} and the subsequent literature relying on this seminal paper. 

In the following we report the full compact expressions of $\textbf{P}(t_{i j}, t_{i j+1})$, $\partial \textbf{P}(t_{i j}, t_{i j+1}) / \partial \theta_w$ and $\partial^2 \textbf{P}(t_{i j}, t_{i j+1}) / \partial \theta_w \partial \theta_{w'}$.
For simplicity, we will drop the dependence on $i$, $j$ and $\boldsymbol\theta$ and define $\delta t = t_{i j+1} - t_{i j}$.

Let $\textbf{Q} = \textbf{A} \boldsymbol{\Gamma} \textbf{A}^{-1}$ be the eigendecomposition of the transition intensity matrix, which is constant over the generic time interval of length $\delta t$, with $\textbf{A}$ the matrix of eigenvectors and $\boldsymbol{\Gamma} = \diag[\gamma_1, \dots, \gamma_C]$ the diagonal matrix containing the eigenvalues. Then
\begin{align}
    &\textbf{P}(\delta t) = \textbf{A} \diag(\exp\big[ \gamma_1 \delta t \big], \dots, \exp\big[ \gamma_Y \delta t  \big]) \textbf{A}^{-1}, \label{eq:P} \\
    &\dfrac{\partial }{\partial \theta_w } \textbf{P}(\delta t)  = \textbf{A} \textbf{U}_{w} \textbf{A}^{-1}, \label{eq:dP} \\ 
    &\dfrac{\partial^2 }{\partial \theta_w \partial \theta_{w'}} \textbf{P}(\delta t) = \textbf{A} ( \check{\textbf{U}}_{w w'} +  \dot{\textbf{U}}_{w w'} + \dot{\textbf{U}}_{w' w} ) \textbf{A}^{-1}, \label{eq:d2P}
\end{align}
where $\textbf{U}_{w} = \mathbf{G}^{(w)} \circ \mathbf{E}$ and $\check{\textbf{U}}_{w w'} = \mathbf{G}^{(w w')} \circ \mathbf{E}$, with $\textbf{E}[l,m] = \frac{\exp[\gamma_l \delta t] - \exp[\gamma_m \delta t]}{\gamma_l - \gamma_m}$ when $\gamma_l \neq \gamma_m$ and $\textbf{E}[l,m] = \delta t e^{\gamma_l \delta t}$ when $\gamma_l = \gamma_m$, $\textbf{G}^{(w)} = \textbf{A}^{-1} \dfrac{\partial \textbf{Q}}{\partial \theta_w} \textbf{A}$, $\textbf{G}^{(w w')} = \textbf{A}^{-1} \dfrac{\partial^2 \textbf{Q}}{\partial \theta_w \partial \theta_{w'}} \textbf{A}$ and 
\begin{align*}
    \dot{\textbf{U}}_{w w'}[l,m] = \sum\limits_{y = 1}^Y  G^{(w)}_{ly} G^{(w')}_{ym}
    \begin{cases}  
    \dfrac{e^{\gamma_l \delta t} - e^{\gamma_y \delta t}}{(\gamma_l - \gamma_y)(\gamma_y - \gamma_m)} - \dfrac{e^{\gamma_l \delta t} - e^{\gamma_m \delta t}}{(\gamma_l - \gamma_m)(\gamma_y - \gamma_m)}, \quad \gamma_l \neq \gamma_y  \neq \gamma_m 
    \vspace{0.2cm} \\ \dfrac{t e^{\gamma_l \delta t}}{\gamma_l - \gamma_m} - \dfrac{e^{\gamma_l \delta t} - e^{\gamma_m \delta t}}{(\gamma_l - \gamma_m)^2}, \quad \gamma_l = \gamma_y  \neq \gamma_m 
    \vspace{0.2cm} \\ \dfrac{e^{\gamma_l \delta t} - e^{\gamma_m \delta t}}{(\gamma_l - \gamma_m)^2} - \dfrac{t e^{\gamma_m \delta t}}{\gamma_l - \gamma_m}, \quad \gamma_m = \gamma_y  \neq \gamma_l
    \vspace{0.2cm} \\ \dfrac{t e^{\gamma_l \delta t}}{\gamma_l - \gamma_y} - \dfrac{e^{\gamma_l \delta t} -  e^{\gamma_y \delta t}}{(\gamma_l - \gamma_y)^2}, \quad \gamma_l = \gamma_m  \neq \gamma_y 
    \vspace{0.2cm} \\ \dfrac{1}{2} \delta t^2 e^{\gamma_l \delta t}, \quad \gamma_l = \gamma_m = \gamma_y
    \end{cases},
\end{align*}
where $G_{lm}^{(w)}$ is the $(l,m)$ element of matrix $\textbf{G}^{(w)}$. $\dot{\textbf{U}}_{w' w}$ is obtained in the same way as $\dot{\textbf{U}}_{w w'}$ but with $w$ and $w'$ swapped wherever they appear. We refer the reader to \cite{kalbfleisch1985analysis} for the proofs of (\ref{eq:P}) and (\ref{eq:dP}) and to \cite{kosorok1995technical} for the proof of (\ref{eq:d2P}).

Note that $\partial \textbf{Q} / \partial \theta_w$ and $\partial^2 \textbf{Q} / \partial \theta_w \partial \theta_{w'}$ are matrices whose $(r,r')$ elements are given, respectively, by $\partial q^{(rr')}(t_{ij}) / \partial \theta_w$ and $\partial^2 q^{(rr')}(t_{ij}) / \partial \theta_w \partial \theta_{w'}$ for $w, w' = 1, \dots, W$. Further, the first derivatives of the transition intensity matrix are already available from the computation of the first derivatives of the transition probabilities, hence only second derivatives have to be computed anew. Matrices $\textbf{A}$, $\textbf{A}^{-1}$ and $\boldsymbol{\Gamma}$ also need to be computed only once, when obtaining matrix \textbf{P}. 



Regarding the implementation of the quantities of interest, the number of operations grows quickly as $n_i$, $N$, $C$ and $W$ increase. Specifically, $\textbf{Q}$ (and its eigendecomposition), $\partial \textbf{Q} / \partial \theta_w$, $ \partial^2 \textbf{Q} / \partial \theta_w \partial \theta_{w'}$, $\textbf{P}$, $\partial \textbf{P} / \partial \theta_w$ and $ \partial^2 \textbf{P} / \partial \theta_w \partial \theta_{w'}$, for $w, w' = 1, \ldots, W$, have to be computed $\sum_{i = 1}^N n_i - N$ times and then combined. To reduce computational cost, the proposed implementation exploited the upper-triangle form of the above mentioned matrices and the presence of structural zero-values in them. We also exploited parallel computing to obtain the log-likelihood, analytical score and information matrix more quickly; the overall run-time of the algorithm can be cut by a factor proportional to the number of cores in the user's computer. 

\section{Inference} \label{sec:inference}


The construction of confidence intervals is based on the results of \cite{Wo2016} for models fitted via penalized log-likelihoods of the general form (\ref{Penloglik}). Specifically, we employ the Bayesian approximation $\boldsymbol\theta \stackrel{\cdot}{\sim} \mathcal{N}(\widehat{\boldsymbol\theta},\V_{\boldsymbol\theta})$, where $\V_{\boldsymbol\theta} = -\textbf{H}_p(\widehat{\boldsymbol\theta})^{-1}$ with $\hat{\boldsymbol{\theta}}$ the estimated model parameter and $\textbf{H}_p(\boldsymbol\theta) = \textbf{H}(\boldsymbol\theta) - \textbf{S}_{\boldsymbol\lambda}$ the penalized Hessian. This approach follows the notion that penalization in estimation implicitly assumes that wiggly models are less likely than smoother ones, which translates into the following prior specification for $\boldsymbol\theta$, $f_{\boldsymbol\theta} \propto \exp\left\{- \boldsymbol\theta\ts \textbf{S}_{\boldsymbol\lambda} \boldsymbol\theta/2\right\}$. Interestingly, using $\V_{\boldsymbol\theta}$ gives close to across-the-function frequentist coverage probabilities because it accounts for both sampling variability and smoothing bias \cite[e.g.,][]{MarraWood}.  

Intervals for linear functions of the model coefficients, e.g. smooth components, are obtained using the result just shown for $\boldsymbol\theta$. For nonlinear functions of the model coefficients, intervals can be conveniently obtained by posterior simulation. For example, $(1-\alpha)100\%$ intervals for the $(r,r')$ transition intensity can be derived as follows:
\begin{enumerate}
    \item Draw $n_{sim}$ random vectors $\bbeta^{(1, rr')}, \dots, \bbeta^{(n_{sim}, rr')}$ from $\mathcal{N}(\widehat{\bbeta^{(rr')}},\V_{\bbeta^{(rr')}})$, where $\widehat{\bbeta^{(rr')}}$ is the estimated model parameter.
    \item Calculate $n_{sim}$ simulated realisations of the quantity of interest, such as $q^{(rr')}(t)$. For fixed $\textbf{x}$ and $t$, one would obtain $\textbf{q}_{sim}^{(rr')} = (q^{(1, rr')}, \ldots, q^{(n_{sim}, rr')})\ts$ using $\bbeta^{(1, rr')}, \dots, \bbeta^{(n_{sim}, rr')}$ respectively.
    \item Using $\textbf{q}_{sim}^{(rr')}$, calculate the lower, $\alpha/2$, and upper, $1-\alpha/2$, quantiles, where $\alpha$ is typically set to $0.05$.
\end{enumerate}

\noindent Values of $n_{sim}$ larger than $100$ do not usually lead to tangible differences in the intervals. Note that the distribution of nonlinear functions of the model parameters need not be symmetric. Intervals for the transition probabilities can be obtained by applying the above procedure to the \textbf{Q} matrices and then deriving the corresponding \textbf{P} matrices, as explained in Section \ref{sec:trans-prob-computation}. 

P-values for the terms in the model are obtained by using the results summarised in \citet[][Section 6.12]{Wood}, which are based on $-\textbf{H}_p(\boldsymbol\theta)^{-1}$. Model building can be aided using tools such as the Akaike information criterion \citep[AIC,][]{akaike1998information} and the Bayesian information criterion \citep[BIC,][]{schwarz1978estimating}. The AIC and BIC are defined as $-2 \boldsymbol\ell(\boldsymbol\theta) + 2 edf$ and $-2 \boldsymbol\ell(\boldsymbol\theta) + \log(\check{n}) edf$, respectively, where the log-likelihood is evaluated at the penalized parameter estimates, $\check{n}$ is the sample size and the effective degrees of freedom are given by $edf = \text{tr}(\textbf{O})$, with $\text{tr}(\cdot)$ the trace function and $\textbf{O} = \sqrt{-\textbf{H}(\boldsymbol\theta)}\left(-\textbf{H}_p(\boldsymbol\theta)\right)^{-1} \sqrt{-\textbf{H}(\boldsymbol\theta)}$ \citep{Marra2019}.


\section{Empirical applications}\label{sec:case-studies}

The next two sections describe and apply the proposed methodology to the two motivating case studies on the onset of CAV and on cognitive decline in the English Longitudinal Study of Ageing (ELSA) population.    

\subsection{CAV study}\label{sec:CAV-study}

\begin{figure}[htb!]
\begin{minipage}{0.5\textwidth}
  \centering
\includegraphics[scale = 0.6]{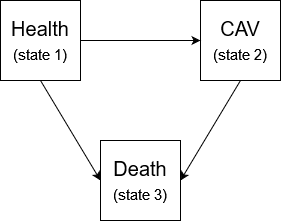}
\captionof{figure}{Diagram of the possible IDM disease trajectories.}\label{fig:IDM}
\end{minipage}
\hfill
\begin{minipage}{0.5\textwidth}
    \centering
    \begin{tabular}{c  c c c}
         & state 1 & state 2 & state 3  \\
      state 1 & 1314 & 223 & 136 \\
      state 2 & 0 & 411 & 105 \\
      state 3 & 0 & 0 & 0
    \end{tabular}
\captionof{table}{Number of observations for each pair of states in the CAV dataset.}\label{tab:CAV_counts}
\end{minipage}
\end{figure}

The heart transplant monitoring data used here are openly accessible from the \texttt{R} package \texttt{msm}. The dataset contains 2846 observations, relating to 622 patients, and is about angiographic (approximately yearly) examinations of heart transplant recipients where the grade of CAV (not present, mild/moderate or severe) is recorded. The additional time event of death is also registered and known exactly (within one day). It follows that the likelihood contributions involved here are those relating to interval censored living states and to exactly observed absorbing states. Available baseline covariates include age of the donor (\texttt{dage}) and primary diagnosis of ischaemic heart disease (IHD, \texttt{pdiag}), which are known to be major risk factors for CAV onset. In line with \cite{machado2021penalised}, we remove eight individuals for which the principal diagnosis is not known and exclude observations which occurred beyond 15 years from the transplant. The resulting dataset contains $\sum_{i = 1}^N n_i = 2803$ observations of $N = 614$ patients. We consider IDMs where the states are (1) health (2) CAV onset (mild/moderate or severe) and (3) death. A diagram representing the process is displayed in Figure \ref{fig:IDM} while Table \ref{tab:CAV_counts} reports the number of observations available for each pair of states in the dataset. Note that the sum of these counts provides the sample size, $\check{n} = 2189$.

The most flexible IDM considered in the literature for this case study is based on equation
\begin{align}\label{eq:mariano_model}
    q^{(rr')}(t_{ij}) = \exp \left[ \beta_{0}^{(rr')} + s_1^{(rr')}(t_{ij}) + \beta_2 \texttt{dage}_{ij} + \beta_3 \texttt{pdiag}_{ij} \right],
\end{align}
for $(r,r') \in \left\{(1,2), (1,3), (2,3) \right\}$, where \textit{t} is the time since transplant, the smooth term is represented by a cubic regression spline with 10 basis functions and second order penalty, and $\beta_2$ and $\beta_3$ are covariate effects which are constrained to be equal across the three transitions, hence the lack of superscript \citep{machado2021penalised}. Model fitting was conducted using the bespoke \texttt{R} code provided by the authors, which took about 3.5 days to reach convergence, on a laptop with Windows 10, Intel 2.20 GHz processor, 16 GB of RAM and eight cores. The resulting AIC was 2931.7. No justification was provided for setting $\beta_2^{(rr')}=\beta_2$ and $\beta_3^{(rr')}=\beta_3$ which may be too restrictive to estimate adequately the effects of \texttt{dage} and \texttt{pdiag}.

The proposed methodology allows us to fit a more general class of models, thanks to the carefully designed estimation framework detailed in this paper. We start by considering a generalisation of (\ref{eq:mariano_model}), where covariate effects are no longer subject to equality constraints, i.e.
\begin{align}\label{ours}
    q^{(rr')}(t_{ij}) = \exp \left[ \beta_{0}^{(rr')} + s_1^{(rr')}(t_{ij}) + \beta_2^{(rr')} \texttt{dage}_{i} + \beta_3^{(rr')} \texttt{pdiag}_{i} \right].
\end{align}
The run-time of \texttt{flexmsm} was 59 minutes, and the resulting AIC equal to 2915.2. Using different spline definitions and increasing $J_1^{(rr')}$ did not lead to tangible empirical differences. Note that employing an approximation of the Hessian, based on first order derivatives, led to convergence failure when fitting the above model as well as those considered at the end of this section. This finding is supported by the simulation study in Supplementary Material \ref{suppl:sim-study-IDM} which shows that basing parameter estimation on an approximate information matrix leads to convergence issues. The study also demonstrates the empirical effectiveness of the proposed approach.

Table \ref{tab:est_effects} reports the effects for \texttt{dage} and \texttt{pdiag}, and their standard errors, resulting from the models based on equations (\ref{eq:mariano_model}) and (\ref{ours}). As the table shows, the constrained coefficients are, roughly speaking, the averages of the respective unconstrained ones. In this case, imposing constraints does not allow one to uncover the differing effects of the risk factors in the different trajectories. Specifically, the results from model (\ref{ours}) indicate that \texttt{dage} and \texttt{pdiag} increase the risks of moving from state 1 to state 2 and from 1 to 3, and that these variables do not play a role in the transition $2 \rightarrow 3$. The curve estimates for the $s_1^{(rr')}(t_{ij})$ (not reported here for the sake of space) were similar across the two models.

\begin{table}[htb!]
    \centering
    \begin{tabular}{c r r c}
    \hline
                              & \texttt{dage} & \texttt{pdiag}  \\
                              \hline
          $1 \rightarrow 2$ &  $ 0.023 \ (0.006)$ & $0.414 \ (0.132)$ \\
          $1 \rightarrow 3$ &  $ 0.040 \ (0.011)$ & $0.341 \ (0.255)$ \\
          $2 \rightarrow 3$ & $-0.016 \ (0.009)$ & $0.002 \ (0.178)$ \\
    \hline
    $1 \rightarrow 2, 1 \rightarrow 3, 2 \rightarrow 3$ & $ 0.018 \ (0.004)$ & $0.274 \ (0.096)$ \\
                              \hline                                  
    \end{tabular}
    \caption{Estimated covariate effects and related standard errors (between brackets) for donor age (\texttt{dage}) and principal diagnosis of IHD (\texttt{pdiag}) obtained using the proposed model fitted by \texttt{flexmsm} (first three lines) and the constrained model of \cite{machado2021penalised} fitted using the related bespoke \texttt{R} code (last line).}
    \label{tab:est_effects}
\end{table}

\begin{figure}[htb!]
\begin{minipage}{0.3\textwidth}
  \centering
\includegraphics[scale = 0.4, angle = 270]{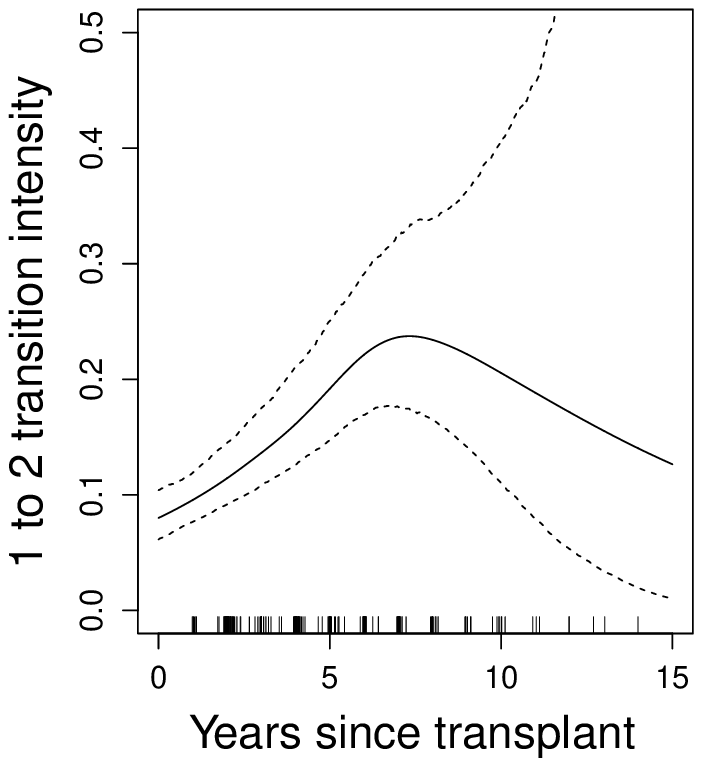}
\end{minipage}
\hfill
\begin{minipage}{0.3\textwidth}
  \centering
\includegraphics[scale = 0.4, angle = 270]{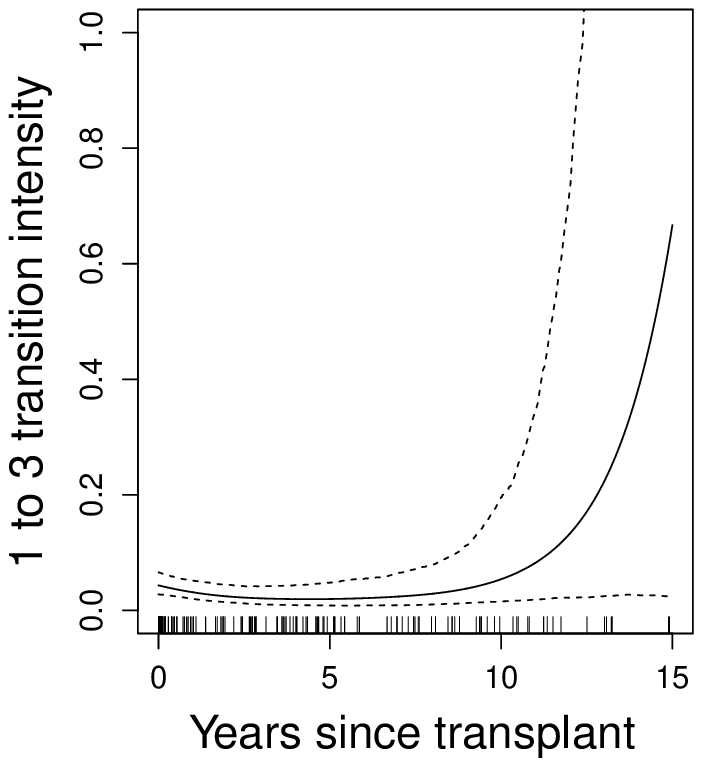}
\end{minipage}
\hfill
\begin{minipage}{0.3\textwidth}
  \centering
\includegraphics[scale = 0.4, angle = 270]{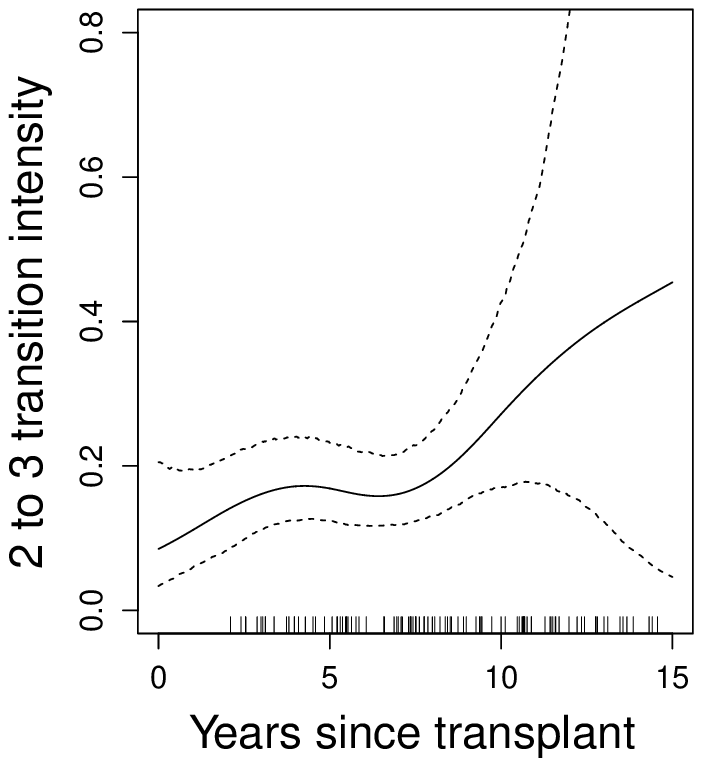}
\end{minipage}
\caption{Estimated transition intensities obtained with \texttt{flexmsm} for $q^{(12)}(\cdot)$, $q^{(13)}(\cdot)$ and $q^{(23)}(\cdot)$ (from left to right) when \texttt{dage = 26} and \texttt{pdiag = 1}, with 95\% intervals derived as detailed in Section \ref{sec:inference}. The `rug plot', at the bottom of each graph, shows the empirical distribution of the transition times. Because we are dealing with an intermittent observation scheme, the time intervals have been represented by plotting the right extremity of each observed interval (the left extremity or mid-point could have been equivalently chosen). Recall that the aim of the rug plot is to highlight regions where the occurrence of a specific transition is rare, hence explaining the width of the intervals across sections.}
\label{fig:CAV_q}
\end{figure}

Figure \ref{fig:CAV_q} shows the estimated transition intensities, and $95\%$ intervals, when \texttt{dage} is equal to 26 years and \texttt{pdiag} is equal to 1 (i.e., the principal diagnosis is IHD). The risk of moving from state 1 to state 2 increases until about 7 years since transplant; after that the situation is uncertain. The risk for the transition $1 \rightarrow 3$ is fairly low and constant until about 10 years, after which it starts increasing steeply. For transition $2 \rightarrow 3$, the risk increases overall. As expected, the intervals are wide when data are scarce. The same exercise can be repeated for different combinations of \texttt{dage} and \texttt{pdiag}. It should be noted that the CAV dataset provided in the \texttt{R} package \texttt{msm} does not indicate the amount of follow-up that occurred after the last angiogram for patients who survived. The stark upward trends exhibited by the estimated intensity functions for the transitions into the death state can thus be explained as an artifact of this. 

Estimated transition intensities provide valuable information about the risks of moving across states. However, interpretation may be more intuitive when transition probabilities are analyzed. Setting, e.g., \texttt{dage = 26} and \texttt{pdiag = 1} and assuming yearly piecewise constant transition intensities, five-year transition probabilities can be obtained by exploiting the Chapman-Kolmogorov equations \citep{cox1977theory}. These allow us to write $\hat{\textbf{P}}(0, 5) = \hat{\textbf{P}}(0, 1) \times \hat{\textbf{P}}(1, 2) \times \dots \times \hat{\textbf{P}}(4, 5)$, where the probabilities over each sub-interval are obtained using the corresponding transition intensity matrix, i.e. $\hat{\textbf{Q}}(t)$, for $t = 0, 1, \dots, 4$, respectively. The resulting estimated transition probability matrix and 95\% intervals (obtained through the method detailed in Section \ref{sec:inference}) are
\begin{align*}
    \hat{\textbf{P}}(0,5) =
    \begin{bmatrix}
        0.48 & (0.42, 0.53) & 0.29 & (0.24, 0.34) & 0.23 & (0.19, 0.29) \\
        0 &      & 0.51 & (0.35, 0.63) & 0.49 & (0.37, 0.64) \\
        0 &      & 0 &           & 1 &  
    \end{bmatrix}.
\end{align*}
For instance, given a healthy starting point, there is a 29\% chance of developing CAV five years after the transplant occurred. Similarly, there is a 23\% chance of dying within the same time frame, given the same starting point.

\begin{figure}[htb!]
\begin{minipage}{0.45\textwidth}
  \centering
\includegraphics[scale = 0.65, angle = 270]{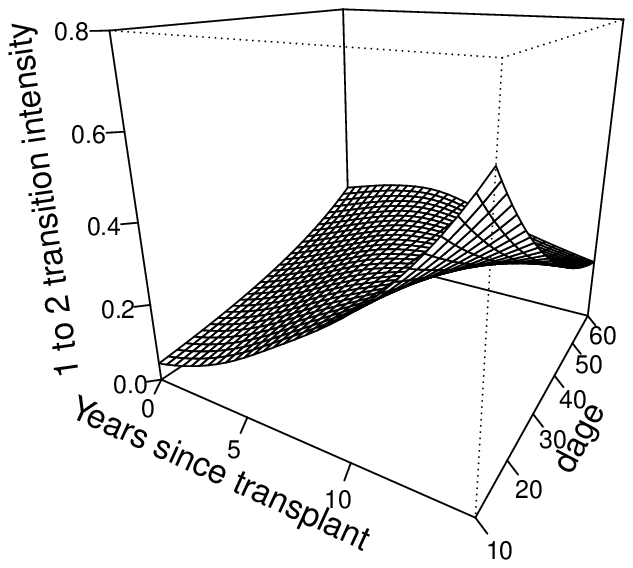}
\end{minipage}
\hfill
\begin{minipage}{0.45\textwidth}
  \centering
\includegraphics[scale = 0.5, angle = 270]{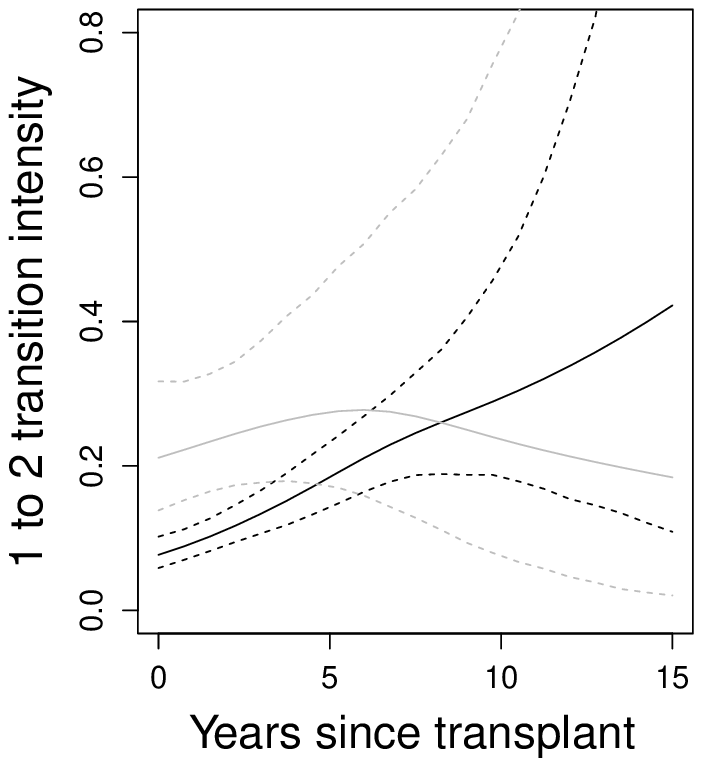}
\end{minipage}
\caption{Left panel: estimated transition intensity surface, obtained with \texttt{flexmsm} when including a time-dependent effect of the donor age. Right panel: sections of the estimated transition intensity surface at $\texttt{dage} = 26$ (black) and $\texttt{dage} = 56$ (grey), along with their respective 95\% confidence intervals (black and grey dashed lines, respectively).}\label{fig:CAV_2Ds}
\end{figure}

We also assessed the possible presence of nonlinear effects of \texttt{dage}. This was achieved by replacing $\beta_2^{(rr')} \texttt{dage}_{ij}$ with $s_2^{(rr')}(\texttt{dage}_{ij})$ in equation (\ref{ours}), where the smooth terms were represented in the same way as for $s_1^{(rr')}(t_{ij})$; the impacts were found to be linear. Finally, we specified a model that allows for interaction terms, i.e.     
\begin{align*}
    q^{(rr')}(t_{ij}) = \exp \left[ \beta_{0}^{(rr')} + s_1^{(rr')}(t_{ij}) + s_2^{(rr')}(\texttt{dage}_{ij}) + s_3^{(rr')}(t_{ij}, \texttt{dage}_{ij}) + \beta_4^{(rr')} \texttt{pdiag}_{ij} \right],
\end{align*}
where $s_3^{(rr')}(t_{ij}, \texttt{dage}_{ij})$ is a tensor product interaction between \texttt{dage} and time whose marginals are cubic regression splines. Here, the main effects and their interaction are modelled separately, thus leading to more flexibility in determining the complexity of the effects \citep[][Section 5.6.3]{Wood}. Let us consider, for instance, the results for transition $1 \rightarrow 2$ shown in Figure \ref{fig:CAV_2Ds}. The left panel reports the estimated transition intensity surface, which is a bivariate function of time and \texttt{dage}. This plot can be read by sectioning the surface, with respect to either of the two arguments, and assessing how the resulting curve varies with respect to the other covariate. The right panel reports two sections of the surface obtained by fixing \texttt{dage} at 26 and 56 years, along with their 95\% confidence intervals. The scarcity of data for the two sections helps explaining the wide confidence intervals, particularly past a certain point. For this reason, it makes more sense to focus on the first few years since the transplant took place. The risk of developing CAV is almost three times higher with a 56 year old donor as compared to a 26 year old donor right after the transplant, and remains overall higher in the following few years. This is in line with the expectation that older donors are associated with higher chances of disease occurrence.

\subsection{ELSA study}\label{sec:ELSA-study}

The \href{https://www.elsa-project.ac.uk/}{ELSA} collects data from people aged over 50 to understand all aspects of ageing in England. More than 18000 people have taken part in the study since it started in 2002, with the same people re-interviewed every two years, hence giving rise to an intermittently observed scheme. ELSA collects information on physical and mental health, wellbeing, finances and attitudes around ageing, and tracks how these change over time. The data can be downloaded from the \href{https://beta.ukdataservice.ac.uk/datacatalogue/series/series?id=200011}{UK Data Service} by registering and accepting an End User License. 

\begin{figure}[htb!]
\begin{minipage}{0.4\textwidth}
  \centering
\includegraphics[scale = 0.22]{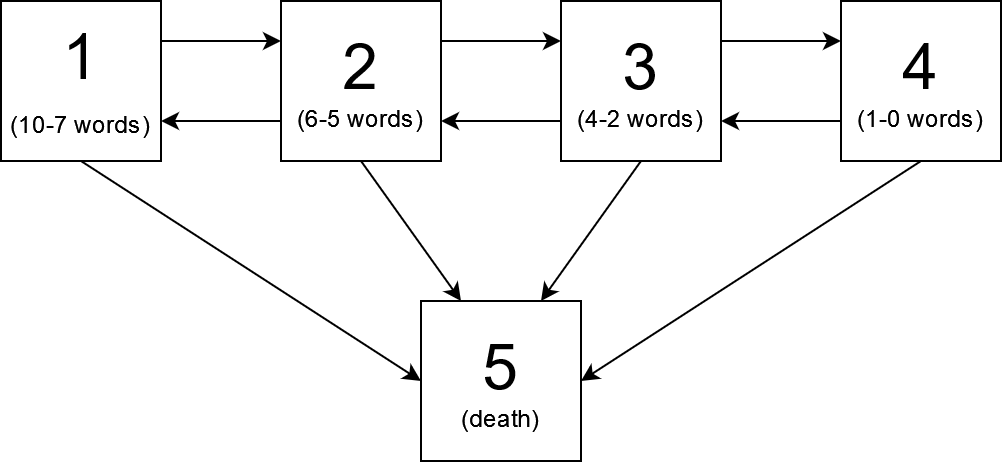}
\captionof{figure}{Diagram of the possible five-state process disease trajectories.}\label{fig:ELSA}
\end{minipage}
\hfill
\begin{minipage}{0.5\textwidth}
    \centering
    \begin{tabular}{c c c c c c}
           & state 1 & state 2 & state 3 & state 4 & state 5  \\
         state 1 & 225 & 194 & 58 & 5 & 11 \\
         state 2 & 209 & 600 & 384 & 54 & 46 \\
         state 3 & 59 & 383 & 732 & 152 & 94 \\
         state 4 & 8 & 42 & 117 & 154 & 70
    \end{tabular}
\captionof{table}{Number of observations for each pair of states in the ELSA dataset.}\label{tab:ELSA-counts}
\end{minipage}
\end{figure}

For this study, interest lies in assessing cognitive function in the older population. This is measured through the score obtained on a test in which participants are asked to remember words in a delayed recall from a list of ten, with the score given by the number of words remembered. In line with \cite{machado2021penalised}, we use a random sample of $N = 1000$ individuals from the full population, leading to 4597 observations, and create four score groups to obtain a five-state process with the fifth state given by the occurrence of death (which is an exactly observed absorbing state). The intermediate states are given by $\{10, 9, 8, 7\}$, $\{6, 5\}$, $\{4, 3, 2\}$ and $\{1, 0\}$ words remembered, respectively. Both forward and backward transitions are allowed between the intermediate states to account for possible improvements or fluctuations through the years in the cognitive function of the participants. In fact, although interest lies mostly in cognitive decline, the opposite trend is also observed as shown in Table \ref{tab:ELSA-counts}. A diagram representing the assumed process is reported in Figure \ref{fig:ELSA}. Further, 221 participants die during the observation period. The time scale is defined by subtracting 49 years to the age of the individuals. 
A potential drawback of this analysis is that the quantity defining the states, i.e. the number of words recalled, is a noisy measure, which has the potential to lead to classification error. Future work will focus on extending the current approach to support hidden Markov models \citep{jackson2003multistate}, which provide a way to handle misclassification such as the one which may arise in this setting.

The most flexible five-state model discussed in the literature for this case study is based on
 \begin{align*}
     q^{(rr')}(t_{ij}) = 
     \begin{cases}
         \exp \left[ \beta_0^{(rr')} + s_1^{(rr')}(t_{ij}) \right] \text{ for } (r,r') \in \{ (1,2), (2,3), (2,5), (3,4), (3,5), (4,5) \} \\
          \exp \left[ \beta_0^{(rr')} \right] \hspace{2.05cm} \text{ for } (r,r') \in \{ (1,5), (2,1), (3,2), (4,3) \}
     \end{cases},
 \end{align*}
where each smooth term is represented by a cubic regression spline with $J_1^{(rr')} = 5$ and second order penalty, with upper bounds for the related smoothing parameters set at $\exp(20)$ \citep{machado2021penalised}. The authors justify the above specification by arguing that the limited information across the age range is probably what causes convergence failures in more general models. No code was made publicly available to replicate the analysis. 

\begin{figure}[htb!]
\begin{minipage}{\textwidth}
  \centering
\includegraphics[scale = 0.6, angle = 270]{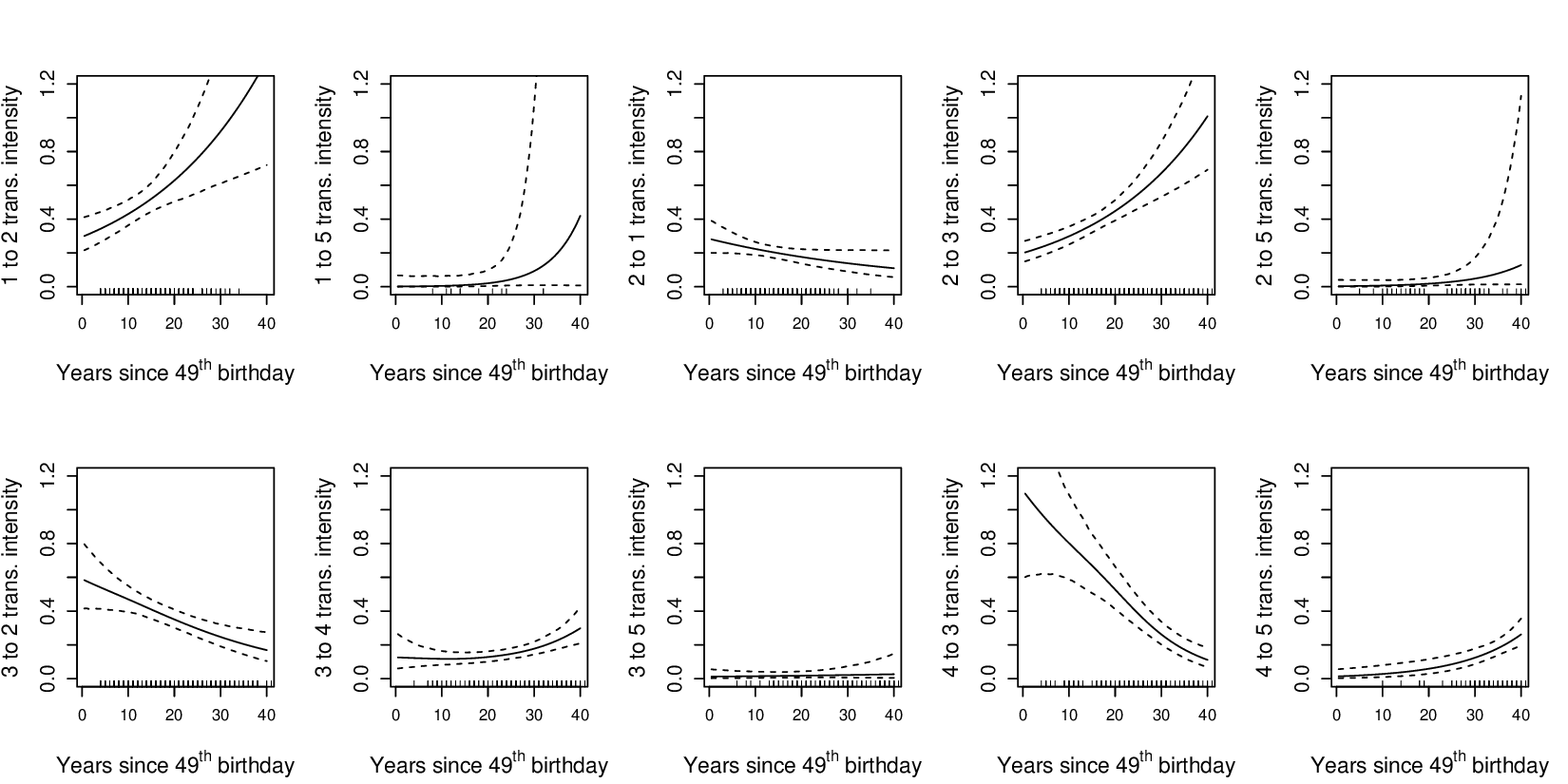}
\end{minipage}
\caption{Estimated transition intensities obtained with \texttt{flexmsm} with the 95\% confidence intervals derived as detailed in Section \ref{sec:inference}.}\label{fig:M3-qhat-plots}
\end{figure}

Using the proposed methodology, we specified
\begin{align}
    q^{(rr')}(t_{ij}) = \exp \left[ \beta_0^{(rr')} + s_1^{(rr')}(t_{ij}) \right] \text{ for } (r,r') \in \mathcal{A},
		\label{gelsa}
\end{align}
with $J_1^{(rr')} = 10$ cubic regression spline bases instead, and $\mathcal{A}$ the set of all the transitions. This model is more general than the one previously considered in that no prior assumptions are made on the way each transition depends on time. As in the CAV study, employing an approximate information matrix led to convergence failure. Figure \ref{fig:M3-qhat-plots} shows the estimated transition intensities, and related 95\% intervals, obtained with \texttt{flexmsm}. As expected, the instantaneous risks of dying are overall smaller than the risks of experiencing further cognitive impairment. As the starting stage reflects more advanced decline, the risk of transitioning to a worse stage becomes a progressively flatter function of time. This shows that once the individuals in the population reach a stage of cognitive impairment, they will typically stay there for the rest of the observation period. Note that there is added value from having modelled the backward transitions through smooth functions of time. For example, we find that the instantaneous chance of improving back to state 3 from a state of cognitive impairment of level 4 decreases considerably faster through time than that of returning to state 1 from state 2. This is in line with expectations as the intermediate stages of cognitive health, i.e. stages 2 and 3, are by far the most frequently observed, with 72\% of the population still alive at the end of the observation period for these categories. The wide 95\% intervals for transitions $1 \rightarrow 5$ and $2 \rightarrow 5$ can be explained by observing, from Table \ref{tab:ELSA-counts}, that these transitions are characterised by the lowest number of observations.


\begin{figure}[htb!]
\begin{minipage}{\textwidth}
  \centering
\includegraphics[scale = 0.60, angle = 270]{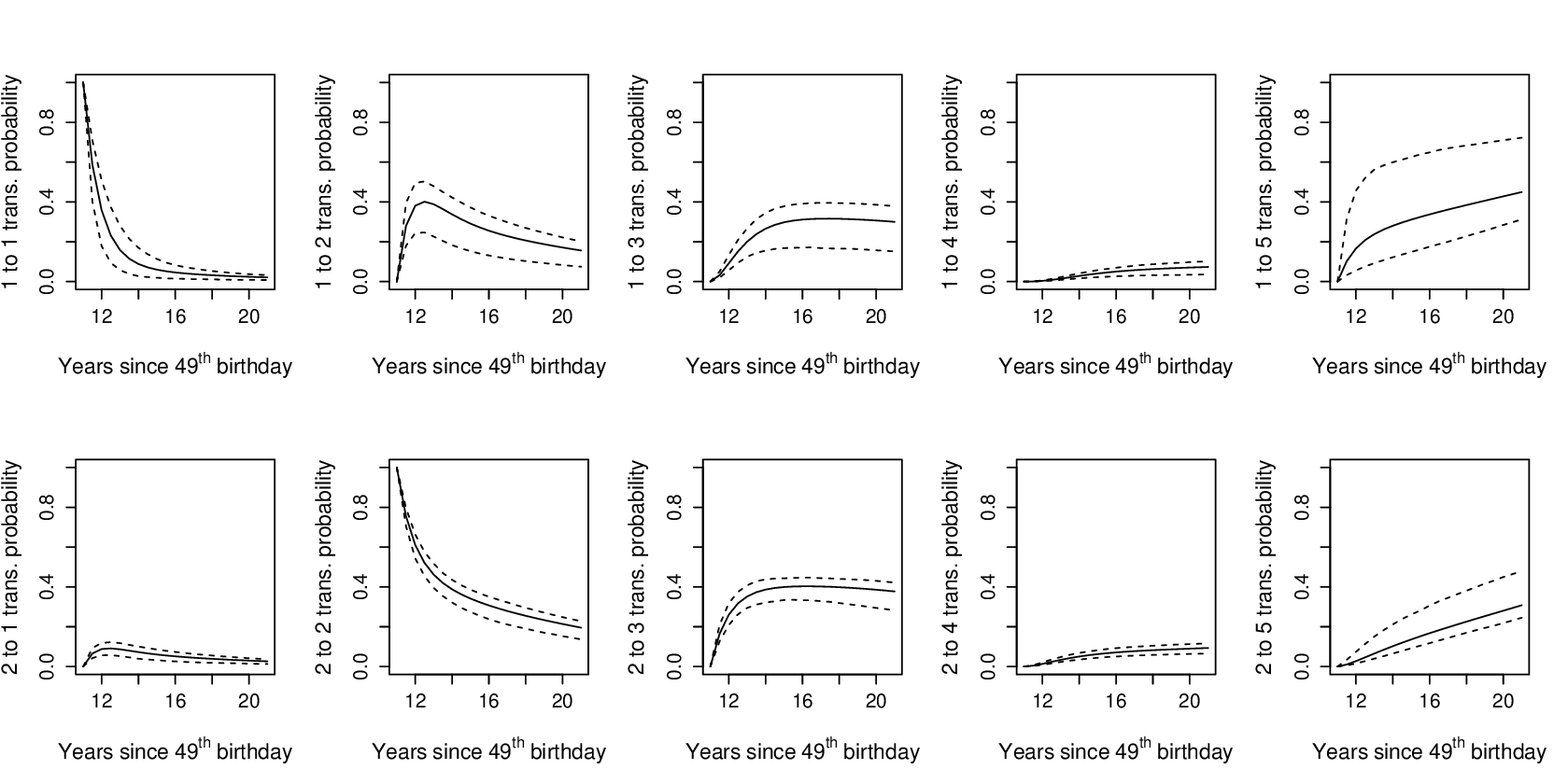}
\end{minipage}
\hfill
\begin{minipage}{\textwidth}
  \centering
\includegraphics[scale = 0.60, angle = 270]{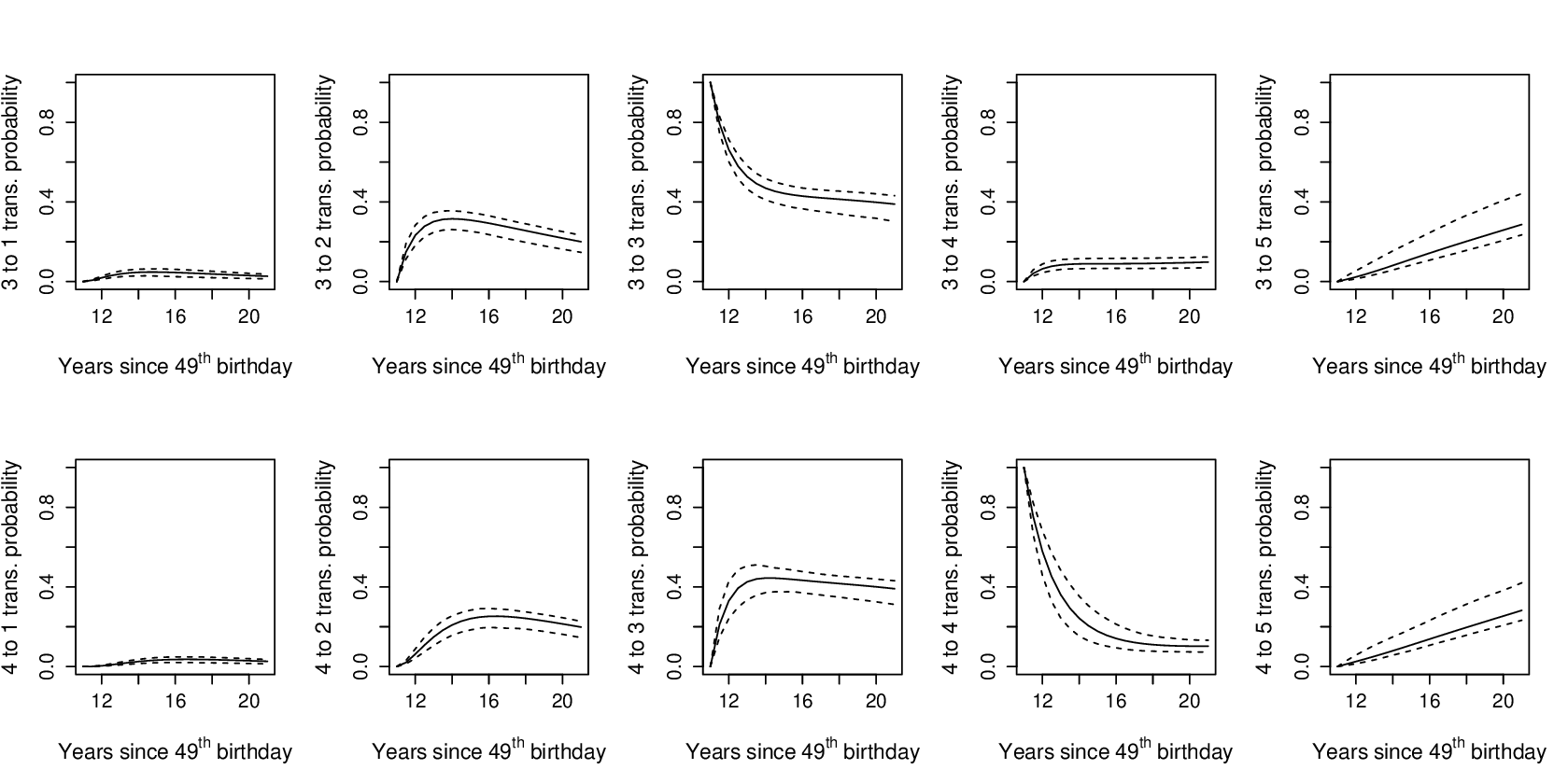}
\end{minipage}
\caption{Transition probabilities for a male individual with less than 10 years of education estimated between 11 and 21 years from their $49^{th}$ birthday, i.e. $\hat{\textbf{P}}(11, \ t)$ and $t \in (11, 21)$. The dashed lines represent the corresponding $95\%$ intervals.}
\label{fig:P-plots-ELSA-edu0}
\end{figure}

We also quantified the effects of two commonly investigated risk factors: \texttt{sex} (0 for male and 1 for female) and \texttt{higherEdu} (0 if the individual has had less than 10 years of education and 1 otherwise) as extracted from the ELSA datasets. This was achieved by simply including $\beta_2^{(rr')} \texttt{sex}_{ij}$ and $\beta_3^{(rr')} \texttt{higherEdu}_{ij}$ in (\ref{gelsa}). We found, for example, that older people with a higher level of education have better memory function, although this does not protect them from cognitive decline as they age \citep[e.g.,][]{cadar2017international}. Overall, the effect of \texttt{sex} was found not to be significant.

Finally, in Figure \ref{fig:P-plots-ELSA-edu0}, we present transition probability plots over 10 years for a 60 year old male with less than 10 years of education. We observe, e.g., that for such individual with stage 2 cognitive health and $\texttt{higherEdu}=0$, the probability of reaching stage 3 by the age of 65 is approximately $40.3\%$, with $95\%$ interval $(33.3\%, 44.7\%)$.


Supplementary Material \ref{suppl:sim-study-five} discusses a simulation study based on a five-state process. The results support the empirical effectiveness of the proposed framework in recovering the true underlying transition intensities in a context which strays from the traditionally explored IDM.

\section{Discussion}\label{sec:discussion}

We propose a general framework for multi-state Markov modeling that allows for different types of processes, with several states and various observation schemes, and that supports time-dependent spline-based transition intensities. This is motivated by the interest in modeling the evolution through time of diseases, with the aim of making statements on their course given specific scenarios or risk factors. The degree of flexibility allowed for the specification of the transition intensities determines the extent to which we can explore and describe the different factors influencing the evolution of a disease. Previous methodological developments have mainly focused on simple parametric forms and time-constant transition intensities, which can be attributed to the lack of an estimation framework capable of supporting more realistic specifications. Attempts addressing this have not been backed by adequate estimation procedures and software implementations.

The key contribution of the paper is the development of an approach that implements and exploits the knowledge of the exact local curvature information. Access to this source of information has allowed us to introduce a modeling framework that has unlocked a host of processes and specifications which were not previously attainable, as demonstrated via the two case studies on cardiac allograft vasculopathy and cognitive decline. To support applicability and reproducibility, we also introduced the \texttt{R} package \texttt{flexmsm}, which is easy and intuitive to use.


Future work will look into improving further the run-time required for model fitting. We are also interested in exploring transformations alternative to the exponential, to enhance the generality allowed by the framework. Note that a Markov process was assumed throughout. Future efforts will look into goodness-of-fit testing \citep[e.g.,][]{titman2009computation} and into the possibility and practical benefits of extending the proposed approach to relax the Markov assumption. Finally, there are circumstances which give rise to multiple dependent multi-state processes, such as the analysis of the evolution of a disease in paired organ systems. In these cases, interest lies in jointly modeling the evolution through time of these events, as the course of one is expected to affect the course of the others. Existing approaches rely on very simple specifications for the marginal processes and restrictive dependence structures among them. The framework proposed in this article will serve as the foundation for the flexible modeling of joint multi-state processes.

\section*{Acknowledgments}

The ELSA dataset was made available through the UK Economic and Social Data Service. The authors of ELSA do not bear any responsibility for the analyses and interpretations presented in this article. 
AE was supported by the UCL Departmental Teaching Assistantship Scholarship. GM and RR were supported by the EPSRC grant EP/T033061/1.

\bibliographystyle{chicago}
\bibliography{main}

\begin{thebibliography}{}

\bibitem[\protect\citeauthoryear{Akaike}{Akaike}{1998}]{akaike1998information}
Akaike, H. (1998).
\newblock Information theory and an extension of the maximum likelihood
  principle.
\newblock In {\em Selected papers of hirotugu akaike}, pp.\  199--213.
  Springer.

\bibitem[\protect\citeauthoryear{Cadar, Robitaille, Clouston, Hofer, Piccinin,
  and Muniz-Terrera}{Cadar et~al.}{2017}]{cadar2017international}
Cadar, D., A.~Robitaille, S.~Clouston, S.~M. Hofer, A.~M. Piccinin, and
  G.~Muniz-Terrera (2017).
\newblock An international evaluation of cognitive reserve and memory changes
  in early old age in 10 european countries.
\newblock {\em Neuroepidemiology\/}~{\em 48\/}(1-2), 9--20.

\bibitem[\protect\citeauthoryear{Cook and Lawless}{Cook and
  Lawless}{2018}]{cook2018multistate}
Cook, R.~J. and J.~F. Lawless (2018).
\newblock {\em Multistate models for the analysis of life history data}.
\newblock CRC Press.

\bibitem[\protect\citeauthoryear{Cox and Miller}{Cox and
  Miller}{1977}]{cox1977theory}
Cox, D.~R. and H.~D. Miller (1977).
\newblock {\em The theory of stochastic processes}, Volume 134.
\newblock CRC press.

\bibitem[\protect\citeauthoryear{Eletti, Marra, and Radice}{Eletti
  et~al.}{2023}]{flexmsm}
Eletti, A., G.~Marra, and R.~Radice (2023).
\newblock {\em flexmsm: A General Framework for Flexible Multi-State Survival
  Modelling}.
\newblock R package version 0.1.0.

\bibitem[\protect\citeauthoryear{Gorfine, Keret, Ben~Arie, Zucker, and
  Hsu}{Gorfine et~al.}{2021}]{gorfine2021marginalized}
Gorfine, M., N.~Keret, A.~Ben~Arie, D.~Zucker, and L.~Hsu (2021).
\newblock Marginalized frailty-based illness-death model: application to the
  uk-biobank survival data.
\newblock {\em Journal of the American Statistical Association\/}~{\em
  116\/}(535), 1155--1167.

\bibitem[\protect\citeauthoryear{Gu, Zeng, Heiss, and Lin}{Gu
  et~al.}{2022}]{gu2022maximum}
Gu, Y., D.~Zeng, G.~Heiss, and D.~Lin (2022).
\newblock Maximum likelihood estimation for semiparametric regression models
  with interval-censored multi-state data.
\newblock {\em arXiv preprint arXiv:2209.07708\/}.

\bibitem[\protect\citeauthoryear{Jackson}{Jackson}{2023}]{msm}
Jackson, C. (2023).
\newblock {\em msm: Multi-State Markov and Hidden Markov Models in Continuous
  Time}.
\newblock R package version 1.7.1.

\bibitem[\protect\citeauthoryear{Jackson et~al.}{Jackson
  et~al.}{2011}]{jackson2011multi}
Jackson, C.~H. et~al. (2011).
\newblock Multi-state models for panel data: the msm package for {R}.
\newblock {\em Journal of statistical software\/}~{\em 38\/}(8), 1--29.

\bibitem[\protect\citeauthoryear{Jackson, Sharples, Thompson, Duffy, and
  Couto}{Jackson et~al.}{2003}]{jackson2003multistate}
Jackson, C.~H., L.~D. Sharples, S.~G. Thompson, S.~W. Duffy, and E.~Couto
  (2003).
\newblock Multistate markov models for disease progression with classification
  error.
\newblock {\em Journal of the Royal Statistical Society Series D: The
  Statistician\/}~{\em 52\/}(2), 193--209.

\bibitem[\protect\citeauthoryear{Joly, Commenges, Helmer, and Letenneur}{Joly
  et~al.}{2002}]{joly2002penalized}
Joly, P., D.~Commenges, C.~Helmer, and L.~Letenneur (2002).
\newblock A penalized likelihood approach for an illness--death model with
  interval-censored data: application to age-specific incidence of dementia.
\newblock {\em Biostatistics\/}~{\em 3\/}(3), 433--443.

\bibitem[\protect\citeauthoryear{Kalbfleisch and Lawless}{Kalbfleisch and
  Lawless}{1985}]{kalbfleisch1985analysis}
Kalbfleisch, J.~D. and J.~F. Lawless (1985).
\newblock The analysis of panel data under a markov assumption.
\newblock {\em Journal of the American Statistical Association\/}~{\em
  80\/}(392), 863--871.

\bibitem[\protect\citeauthoryear{Kosorok and Chao}{Kosorok and
  Chao}{1995}]{kosorok1995technical}
Kosorok, M.~R. and W.-H. Chao (1995).
\newblock Further details on the analysis of longitudinal ordinal response data
  in continuous time.
\newblock Technical Report~92, University of Wisconsin, Madison, Dept. of
  Biostatistics.

\bibitem[\protect\citeauthoryear{Kosorok and Chao}{Kosorok and
  Chao}{1996}]{kosorok1996analysis}
Kosorok, M.~R. and W.-H. Chao (1996).
\newblock The analysis of longitudinal ordinal response data in continuous
  time.
\newblock {\em Journal of the American Statistical Association\/}~{\em
  91\/}(434), 807--817.

\bibitem[\protect\citeauthoryear{Machado, Van~den Hout, and Marra}{Machado
  et~al.}{2021}]{machado2021penalised}
Machado, R. J.~M., A.~Van~den Hout, and G.~Marra (2021).
\newblock Penalised maximum likelihood estimation in multi-state models for
  interval-censored data.
\newblock {\em Computational Statistics \& Data Analysis\/}~{\em 153}, 107057.

\bibitem[\protect\citeauthoryear{Marra and Radice}{Marra and
  Radice}{2020}]{Marra2019}
Marra, G. and R.~Radice (2020).
\newblock Copula link-based additive models for right-censored event time data.
\newblock {\em Journal of the American Statistical Association\/}~{\em
  115\/}(530), 886--895.

\bibitem[\protect\citeauthoryear{Marra and Wood}{Marra and
  Wood}{2012}]{MarraWood}
Marra, G. and N.~S. Wood (2012).
\newblock Coverage properties of confidence intervals for generalized additive
  model components.
\newblock {\em Scandinavian Journal of Statistics\/}~{\em 39\/}(1), 53--74.

\bibitem[\protect\citeauthoryear{Mor{\'e} and Sorensen}{Mor{\'e} and
  Sorensen}{1983}]{more1983computing}
Mor{\'e}, J.~J. and D.~C. Sorensen (1983).
\newblock Computing a trust region step.
\newblock {\em SIAM Journal on Scientific and Statistical Computing\/}~{\em
  4\/}(3), 553--572.

\bibitem[\protect\citeauthoryear{Nocedal and Wright}{Nocedal and
  Wright}{2006}]{Nocedal}
Nocedal, J. and S.~J. Wright (2006).
\newblock {\em Numerical Optimization}.
\newblock Springer-Verlag, New York.

\bibitem[\protect\citeauthoryear{Sabath{\'e}, Andersen, Helmer, Gerds,
  Jacqmin-Gadda, and Joly}{Sabath{\'e} et~al.}{2020}]{sabathe2020regression}
Sabath{\'e}, C., P.~K. Andersen, C.~Helmer, T.~A. Gerds, H.~Jacqmin-Gadda, and
  P.~Joly (2020).
\newblock Regression analysis in an illness-death model with interval-censored
  data: A pseudo-value approach.
\newblock {\em Statistical methods in medical research\/}~{\em 29\/}(3),
  752--764.

\bibitem[\protect\citeauthoryear{Schwarz}{Schwarz}{1978}]{schwarz1978estimating}
Schwarz, G. (1978).
\newblock Estimating the dimension of a model.
\newblock {\em The annals of statistics\/}, 461--464.

\bibitem[\protect\citeauthoryear{Titman}{Titman}{2023}]{nhm}
Titman, A. (2023).
\newblock {\em nhm: Non-Homogeneous Markov and Hidden Markov Multistate
  Models}.
\newblock R package version 0.1.1.

\bibitem[\protect\citeauthoryear{Titman}{Titman}{2009}]{titman2009computation}
Titman, A.~C. (2009).
\newblock Computation of the asymptotic null distribution of goodness-of-fit
  tests for multi-state models.
\newblock {\em Lifetime Data Analysis\/}~{\em 15\/}(4), 519--533.

\bibitem[\protect\citeauthoryear{Titman}{Titman}{2011}]{titman2011flexible}
Titman, A.~C. (2011).
\newblock Flexible nonhomogeneous markov models for panel observed data.
\newblock {\em Biometrics\/}~{\em 67\/}(3), 780--787.

\bibitem[\protect\citeauthoryear{Van Den~Hout}{Van
  Den~Hout}{2017}]{van2016multi}
Van Den~Hout, A. (2017).
\newblock {\em Multi-state survival models for interval-censored data}.
\newblock CRC Press.

\bibitem[\protect\citeauthoryear{Van~den Hout and Matthews}{Van~den Hout and
  Matthews}{2008}]{van2008multi}
Van~den Hout, A. and F.~E. Matthews (2008).
\newblock Multi-state analysis of cognitive ability data: a piecewise-constant
  model and a weibull model.
\newblock {\em Statistics in Medicine\/}~{\em 27\/}(26), 5440--5455.

\bibitem[\protect\citeauthoryear{Williams, Storlie, Therneau, Jr, and
  Hannig}{Williams et~al.}{2020}]{williams2020bayesian}
Williams, J.~P., C.~B. Storlie, T.~M. Therneau, C.~R.~J. Jr, and J.~Hannig
  (2020).
\newblock A bayesian approach to multistate hidden markov models: application
  to dementia progression.
\newblock {\em Journal of the American Statistical Association\/}~{\em
  115\/}(529), 16--31.

\bibitem[\protect\citeauthoryear{Wood}{Wood}{2004}]{Wood04}
Wood, S.~N. (2004).
\newblock Stable and efficient multiple smoothing parameter estimation for
  generalized additive models.
\newblock {\em Journal of the American Statistical Association\/}~{\em
  99\/}(467), 673--686.

\bibitem[\protect\citeauthoryear{Wood}{Wood}{2017}]{Wood}
Wood, S.~N. (2017).
\newblock {\em Generalized Additive Models: An Introduction With R}.
\newblock Second Edition, Chapman \& Hall/CRC, London.

\bibitem[\protect\citeauthoryear{Wood and Fasiolo}{Wood and
  Fasiolo}{2017}]{wood2017generalized}
Wood, S.~N. and M.~Fasiolo (2017).
\newblock A generalized fellner-schall method for smoothing parameter
  optimization with application to tweedie location, scale and shape models.
\newblock {\em Biometrics\/}~{\em 73\/}(4), 1071--1081.

\bibitem[\protect\citeauthoryear{Wood, Pya, and S{\"a}fken}{Wood
  et~al.}{2016}]{Wo2016}
Wood, S.~N., N.~Pya, and B.~S{\"a}fken (2016).
\newblock Smoothing parameter and model selection for general smooth models.
\newblock {\em Journal of the American Statistical Association\/}~{\em
  111\/}(516), 1548--1563.

\bibitem[\protect\citeauthoryear{Yiu, Farewell, and Tom}{Yiu
  et~al.}{2017}]{yiu2017exploring}
Yiu, S., V.~T. Farewell, and B.~D. Tom (2017).
\newblock Exploring the existence of a stayer population with mover--stayer
  counting process models: application to joint damage in psoriatic arthritis.
\newblock {\em Journal of the Royal Statistical Society. Series C, Applied
  Statistics\/}~{\em 66\/}(4), 669.

\end{thebibliography}

\newpage

\appendix

\setcounter{page}{1}

\LARGE
\begin{center}
\textbf{Supplementary Material: "Spline-Based Multi-State Models for Analyzing Disease Progression"}
\end{center}

\normalsize

\section{Log-likelihood contributions}\label{suppl:log-lik-grad-Hess}

This section follows \cite{jackson2011multi}. For a time-inhomogeneous Markov process, the likelihood contribution for the $j^{th}$ observation of the $i^{th}$ unit can take any of the following forms
\begin{align*}
    L_{ij}(\boldsymbol{\theta}) = 
    \begin{cases}
   p^{(z_{i j-1} z_{i j})}(t_{i j-1}, t_{i j}), \quad &\text{if $z_{i j}$ is an interval censored state} \vspace{0.2cm} \\ 
    \exp\bigg[ \int\limits_{t_{i j-1}}^{t_{ij}} q^{(z_{i j-1} z_{i j-1})}(u) du \bigg] q^{(z_{i j-1} z_{i j})}(t_{i j}), \quad &\text{if $z_{i j}$ is an exactly observed state} \vspace{0.2cm} \\ 
    \sum\limits_{c \in \tilde{\mathcal{S}} \subset \mathcal{S}} p^{(z_{i j-1} c)}(t_{i j-1}, t_{i j}), \quad &\text{if $z_{i j}$ is a censored state} \vspace{0.2cm} \\
    \sum\limits_{\substack{c = 1 \\ c \neq z_{i j}}}^C p^{(z_{i j-1} c)}(t_{i j-1}, t_{i j}) q^{(c z_{i j})}(t_{i j}), \quad &\text{if $z_{i j}$ is an exactly observed death state} 
    \end{cases}.
\end{align*}
for $i = 1, \dots, N$, $j = 1, \dots, n_i$, with $N$ the total number of statistical units and $n_i$ the number of observations for unit $i$ and where $p^{(z_{i j-1} z_{i j})}(t_{i j-1}, t_{i j}) = P(Z(t_{ij}) = z_{i j} \mid Z(t_{i j-1}) = z_{i j-1})$. In other words, each pair of consecutively observed states contributes one term to the likelihood. Specifically, if a transition between two transient states is observed and the transition time is interval-censored then the contribution is 
\begin{align*}
    L_{ij}(\boldsymbol{\theta}) = P(Z(t_{ij}) = z_{ij} \mid Z(t_{ij-1}) = z_{ij-1}), 
\end{align*}
If, due to the nature of the process, the transitions to some states are exactly observed, the contribution is
\begin{align*}
    L_{ij}(\boldsymbol{\theta}) = \exp\bigg[ \int\limits_{t_{i j-1}}^{t_{ij}} q^{(z_{i j-1} z_{i j-1})}(u) du \bigg] q^{(z_{i j-1} z_{i j})}(t_{i j}),
\end{align*}
since the process is known to have stayed in state $z_{ij-1}$ between $t_{i j-1}$ and $t_{ij}$ and then jumped from state $z_{i j-1}$ to state $z_{i j}$ at exactly $t_{ij}$. The first term can be explained by observing that
\begin{align*}
    \exp\bigg[ \int\limits_{t_{i j-1}}^{t_{ij}} q^{(z_{i j-1} z_{i j-1})}(u) du \bigg] & = \exp\bigg[ - \int_{t_{i j-1}}^{t_{ij}} \sum\limits_{c \neq z_{i j-1}} q^{(z_{i j-1} c)}(u) du \bigg] \\
    & = \prod\limits_{c \neq z_{i j-1}} \dfrac{\exp\bigg[ - \int_{0}^{t_{ij}} q^{(z_{i j-1} c)}(u) du \bigg] }{\exp\bigg[ - \int_{0}^{t_{i j-1}} q^{(z_{i j-1} c)}(u) du \bigg] },
\end{align*}
which implies that no transition exiting state $z_{i j-1}$ has occurred at time $t_{i j}$ given that it had not occurred by time $t_{i j-1}$ either.

If the state occupied at a given time is only known to lie in a subset $\tilde{\mathcal{S}}$, then it is said to be censored. In this case, the contribution to the likelihood has to account for all the possible trajectories that may have occurred from the last known occupied state to the current observation time. Therefore, the sum over the various probabilities is taken, which will be null if the transition is not allowed. In particular,
\begin{align*}
    L_{ij}(\boldsymbol{\theta}) = \sum\limits_{c \in \tilde{\mathcal{S}} \subset \mathcal{S}} P(Z(t_{i j}) = c \mid Z(t_{i j-1}) = z_{i j-1}).
 \end{align*}

Finally, if the last observed state is a death state, then the time at which the transition occurred is generally assumed to be known. In this case, one needs to account for the possibility that the state occupied before the absorbing state is unknown and thus the contribution to the likelihood is summed over the possible states occupied by the process. The information of the exact observation time $t_{i n_i}$ is included through the transition intensity computed in that time. Here, we have
\begin{align*}
     L_{ij}(\boldsymbol{\theta}) =  \sum\limits_{\substack{c = 1 \\ c \neq z_{i j}}}^C p^{(z_{i j-1} c)}(t_{i j-1}, t_{i j}) q^{(c z_{i j})}(t_{i j}).
\end{align*}

\newpage

\section{\texttt{R} package \texttt{flexmsm}}\label{suppl:code}

To support applicability and reproducibility, the proposed modeling framework has been implemented in the \texttt{R} package \texttt{flexmsm}. The package is straightforward to use, especially if the user is already familiar with the syntax of generalized linear models and generalized additive models (GAMs) in \texttt{R}. The key function is \texttt{fmsm()}, which carries out model fitting and inference, and is exemplified with some of its main arguments in the following code snippet
\begin{verbatim}
out <- fmsm(formula = formula, data = df, 
            id = ID, state = state, 
            death = TRUE, living.exact = NULL, cens.state = -99,
            sp.method = 'perf',
            constraint = NULL, parallel = TRUE, ...)
\end{verbatim}
where the user specifies the model through the argument \texttt{formula} as a \texttt{list()} containing the model specifications for the transition intensities, and the dataset has to be provided through the argument \texttt{data}. This will always have at least three columns: the state column (whose name is provided through the argument \texttt{state}), the column containing the unique IDs (whose name is provided through the argument \texttt{id}) identifying each individual, and a column containing the (intermittent) observation times. The arguments \texttt{death}, \texttt{living.exact} and \texttt{cens.state} allow the user to specify the observation type. If the last state in the process is an exactly observed death state then the user must specify \texttt{death = TRUE}; if there are exactly observed living states then the dataset must contain an additional column with \texttt{TRUE} (or \texttt{1}) if the data point is exactly observed and \texttt{FALSE} (or \texttt{0}) otherwise; the name of this column must be passed through the argument \texttt{living.exact}, which defaults to \texttt{NULL}. If there are any censored states then the user must specify the code used to indicate this through the argument \texttt{cens.state}, which defaults to -99. The \texttt{sp.method} argument specifies the method employed for multiple smoothing parameter estimation (this can be set to \texttt{'perf'} or \texttt{'efs'}). The argument \texttt{constraint} allows the user to specify equality constraints on the covariates. The \texttt{parallel} argument allows the user to exploit parallel computing, in Windows, for the likelihood, gradient and Hessian, thus cutting the run-time of the algorithm by factor proportional to the number of cores on the computer. 



The \texttt{formula} is a \texttt{list()} object whose elements are the off-diagonal elements of the transition intensity matrix. The order of the elements is that given by reading the $\textbf{Q}$ matrix from the first row to the last and from left to right. The equation corresponding to each non-zero transition intensity has to be specified with syntax similar to that used for GAMs, with the response given by the time-to-event variable. Trivially, zero elements have to be specified with a \texttt{0}. For instance, we may consider the following model, with a smooth effect of time $t$ and two covariates $x_1$ and $x_2$, one included linearly and the other as a time-dependant flexible effect, for a transition $r \rightarrow r'$
\begin{align*}
    q^{(rr')}(t_{ij}) = \exp \left[ \beta_0^{(rr')} + s_1^{(rr')}(t_{ij}) + \beta_2^{(rr')} x_{1 ij} + s_3^{(rr')}(x_{2 ij}) + s_4^{(rr')}(t_{ij}, x_{2 ij}) \right].
\end{align*}

This will be specified, in the correct position, as part of the list
\begin{verbatim}
formula <- list(...,
               t ~ s(t) + x1 + s(x2), ti(t, x2), # r -> r' trans.
               ...) 
\end{verbatim}
where $\texttt{\dots}$ represent other possible transition-specific equations or \texttt{0}s for transitions not allowed by the process. The model specified here is only an example and many types of effects are supported. For instance, as the above example shows, time-dependent effects are modelled by using a tensor interaction function $\texttt{ti()}$ on the covariate of interest and time.

Functions \texttt{summary()} and \texttt{plot()} can be used in the usual way to obtain post-estimation summaries for each non-zero transition intensity and the plots of the smooths. In the example above there is a two-dimensional spline, thus \texttt{plot()} will also automatically produce a three-dimensional plot of the surface representing this time-dependent effect. 

Function \texttt{conv.check()} allows the user to check the convergence of the fitted model by providing information on whether the gradient is zero and the Hessian is positive definite. It also provides information on the values taken by the $\textbf{Q}$ matrix since, in practice, we have found that particularly large values are red flags for ill-defined problems, for instance. 

Prediction and plotting of the $\textbf{P}$ and the $\textbf{Q}$ matrices can be carried out through the functions \texttt{P.pred()} and \texttt{Q.pred()}, respectively. For instance, the specification
\begin{verbatim}
     P.hat <- P.pred(out, newdata = newdata, plot.P = TRUE
                     get.CI = TRUE, prob.lev = 0.05)
\end{verbatim}
will provide an object $\texttt{P.hat}$ containing the estimated transition probability matrix corresponding to the time interval and profile of interest, specified through argument \texttt{newdata}. The intermediate transition probabilities corresponding to each sub-interval specified in \texttt{newdata} are also provided. The $100(1-\texttt{prob.lev})\%$ confidence intervals can be obtained by setting \texttt{get.CI = TRUE}. When $\texttt{plot.P = TRUE}$ the transition probabilities are also plotted as function of time over the interval considered, otherwise the plots are suppressed. The analogous output can be obtained for the $\textbf{Q}$ matrix through function \texttt{Q.pred()} with similar syntax. 

To exemplify the usage of the software, we report the code used to fit the models presented in Section \ref{sec:case-studies}. We recall that the IDM specified in Section \ref{sec:CAV-study} is given by
\begin{align*}
    q^{(rr')}(t_{ij}) = \exp \left[ \beta_{0}^{(rr')} + s_1^{(rr')}(t_{ij}) + \beta_2^{(rr')} \texttt{dage}_{ij} + \beta_3^{(rr')} \texttt{pdiag}_{ij}\right].
\end{align*}
This can be fitted in the following way:
\begin{verbatim}
formula <- list(t ~ s(t, bs = 'cr', k = 10) + dage + pdiag, # 1-2
                t ~ s(t, bs = 'cr', k = 10) + dage + pdiag, # 1-3
                0,                                          # 2-1
                t ~ s(t, bs = 'cr', k = 10) + dage + pdiag, # 2-3
                0,                                          # 3-1
                0)                                          # 3-2

fmsm.out <- fmsm(formula = formula, data = Data, 
              id = PTNUM, state = state, death = TRUE,
              sp.method = 'perf', parallel = TRUE)
\end{verbatim}
Here \texttt{bs = 'cr'} and \texttt{k = 10} imply that the smooths of time are specified through cubic regression splines with ten basis functions. We will omit this in the following to avoid redundancies. To obtain the two-dimensional spline based model, it suffices to swap the \texttt{formula} reported above with the following
\begin{verbatim}
formula <- list(t ~ s(t) + s(dage) + ti(t, dage) + pdiag, # 1-2
                t ~ s(t) + s(dage) + ti(t, dage) + pdiag, # 1-3
                0,                                        # 2-1
                t ~ s(t) + s(dage) + ti(t, dage) + pdiag, # 2-3
                0,                                        # 3-1
                0)                                        # 3-2
\end{verbatim}
For the five-state model described in Section \ref{sec:ELSA-study}, the first model explored was
\begin{align*}
    q^{(rr')}(t_{ij}) = \exp \left[ \beta_0^{(rr')} + s_1^{(rr')}(t_{ij}) \right].
\end{align*}
This can be implemented in the following way:
\begin{verbatim}
formula <- list(t ~ s(t) + sex + edu, # 1-2
                0,                    # 1-3
                0,                    # 1-4
                t ~ s(t) + sex + edu, # 1-5
                t ~ s(t) + sex + edu, # 2-1
                t ~ s(t) + sex + edu, # 2-3
                0,                    # 2-4
                t ~ s(t) + sex + edu, # 2-5
                0,                    # 3-1
                t ~ s(t) + sex + edu, # 3-2
                t ~ s(t) + sex + edu, # 3-4
                t ~ s(t) + sex + edu, # 3-5
                0,                    # 4-1
                0,                    # 4-2
                t ~ s(t) + sex + edu, # 4-3
                t ~ s(t) + sex + edu, # 4-5
                0,                    # 5-1
                0,                    # 5-2
                0,                    # 5-3
                0)                    # 5-4

fmsm.out <- fmsm(formula = formula, data = ELSA.df, 
              id = idauniq, state = state, death = TRUE,
              sp.method = 'efs')
\end{verbatim}

\newpage

\section{Parameter estimation}\label{suppl:algorithm}

The algorithm employed for model fitting is characterized by two steps. In the first step, $\blambda$ is held fixed at a vector of values and for a given $\boldsymbol\theta^{[a]}$, where $a$ is an iteration index, equation (\ref{Penloglik}) is maximized using

\begin{equation}
\boldsymbol\theta^{[a+1]} = \boldsymbol\theta^{[a]} + \underset{\textbf{e}:\|\textbf{e}\|\leq \Delta^{[a]}}{\operatorname{arg \ min}} \ \ \breve{\ell_p}(\boldsymbol\theta^{[a]}), 
\label{eq1st1}
\end{equation}
where $\breve{\ell_p}(\boldsymbol\theta^{[a]})=-\left\{\ell_p(\boldsymbol\theta^{[a]})+\textbf{e}\ts\textbf{g}_p(\boldsymbol\theta^{[a]})+\frac{1}{2}\textbf{e}\ts\textbf{H}_p(\boldsymbol\theta^{[a]})\textbf{e}\right\}$, $\textbf{g}_p(\boldsymbol\theta^{[a]})=\textbf{g}(\boldsymbol\theta^{[a]})- \textbf{S}_{\boldsymbol\lambda} \boldsymbol\theta^{[a]}$, and $\textbf{H}_p(\boldsymbol\theta^{[a]}) = \textbf{H}(\boldsymbol\theta^{[a]}) - \textbf{S}_{\boldsymbol\lambda}$. $\textbf{g}(\boldsymbol\theta^{[a]}) = \partial \ell(\boldsymbol\theta) /\partial \boldsymbol\theta|_{\boldsymbol\theta = \boldsymbol\theta^{[a]}}$ and $\textbf{H}(\boldsymbol\theta^{[a]}) = \partial^2 \ell(\boldsymbol\theta)/\partial \boldsymbol\theta \partial \boldsymbol\theta\ts|_{\boldsymbol\theta = {\boldsymbol\theta}^{[a]}}$ are given in Section \ref{sec:par-estimation}, $\|\cdot\|$ denotes the Euclidean norm, and $\Delta^{[a]}$ is the radius of the trust region which is adjusted through the iterations. The first line of (\ref{eq1st1}) uses a quadratic approximation of $-\ell_p$ about $\boldsymbol\theta^{[a]}$ (the so-called model function) to choose the best $\textbf{e}^{[a+1]}$ within the ball centered in $\boldsymbol\theta^{[a]}$ of radius $\Delta^{[a]}$, the trust-region. Throughout the iterations, a proposed solution is accepted or rejected and the trust region adjusted (i.e., expanded or shrunken) based on the ratio between the improvement in the objective function when going from $\boldsymbol\theta^{[a]}$ to $\boldsymbol\theta^{[a+1]}$ and that predicted by the approximation. The use of the observed information matrix gives global convergence guarantees due to \citet{more1983computing}. Importantly, convergence to a point satisfying the second-order sufficient conditions (i.e., a local strict minimiser) is super-linear. Near the solution, the algorithm proposals become asymptotically similar to Newton-Raphson steps, hence benefitting from the resulting fast convergence rate. Trust region algorithms are also generally more stable and faster compared to in-line search methods. See \citet[Chapter 4,][]{Nocedal} for proofs and further details. The starting values $\boldsymbol{\theta}^{[0]}$ are set automatically to small positive values, except for the transition-specific intercepts which are given by the maximum likelihood estimates one would obtain when assuming that the data represent the exact transition times of the corresponding covariate-free time-homogeneous Markov process. Vector $\boldsymbol{\theta}^{[0]}$ can, alternatively, be provided by the user. Importantly, through extensive experimentation, we have found that the algorithm is not particularly sensitive to the choice of starting values.


In the second step, at $\boldsymbol\theta^{[a+1]}$, there are two options to estimate the smoothing parameter vector: the stable and efficient multiple smoothing parameter approach adopted by \cite{Marra2019}, and the generalized Fellner-Schall method of \cite{wood2017generalized}. Both techniques can be employed for fitting penalized likelihood-based models, and require the availability of the analytical score and information matrix. In the former, the following problem is solved
\begin{align}\label{PIRLS1}
    \blambda^{[a+1]}=\underset{\blambda}{\operatorname{arg \ min}} \ \| \textbf{M}^{[a+1]} - \textbf{O}^{[a+1]}\textbf{M}^{[a+1]} \|^2 - \check{n} + 2\text{tr}(\textbf{O}^{[a+1]}).
\end{align}
The idea is to estimate $\blambda$ so that the complexity of the smooth terms not supported by the data is suppressed. This is formalized as $\E \left( \| \muz - \widehat{\bm\mu}_{\textbf{M}} \|^2 \right) = \E\left( \| \textbf{M} - \textbf{O} \textbf{M} \|^2\right) - \check{n} + 2\text{tr}(\textbf{O})$, where $\textbf{M} =\muz + \eb$, $\muz=\sqrt{-\textbf{H}(\boldsymbol\theta)}\boldsymbol\theta$, $\eb = \sqrt{-\textbf{H}(\boldsymbol\theta)}^{-1}\textbf{g}(\boldsymbol\theta)$, $\textbf{O} = \sqrt{-\textbf{H}(\boldsymbol\theta)}\left(-\textbf{H}(\boldsymbol\theta) + \textbf{S}_{\boldsymbol\lambda} \right)^{-1} \sqrt{-\textbf{H}(\boldsymbol\theta)}$, and $\text{tr}(\textbf{O})$ is defined in Section 5 of the main paper. It can be proved that (\ref{PIRLS1}) is approximately equivalent to the AIC with number of parameters given by $\text{tr}(\textbf{O})$. Iteration (\ref{PIRLS1}) is implemented via the routine by \citet{Wood04}, which is based on the Newton method and can evaluate in an efficient and stable manner the terms in (\ref{PIRLS1}), their scores and Hessians, with respect to $\log(\blambda)$.

The approach proposed in \cite{wood2017generalized} is based on a different principle. The starting point is the well established stance that smoothing penalties can be viewed as resulting from improper Gaussian prior distributions on the spline coefficients. This is also the Bayesian viewpoint taken for the inferential result discussed in Section \ref{sec:inference}, and implies the following improper joint log-density, where the dependence on the smoothing parameter has been made explicit,
\begin{align*}
    \log L(\boldsymbol\theta; \boldsymbol\lambda) = \ell(\boldsymbol\theta) - \dfrac{1}{2} \boldsymbol\theta\ts \textbf{S}_{\boldsymbol\lambda} \boldsymbol\theta + \dfrac{1}{2} \log |\textbf{S}_{\boldsymbol\lambda}|.
\end{align*}
The idea is to develop an update for $\boldsymbol\lambda$ that maximises the restricted marginal likelihood $L(\boldsymbol\lambda)$, obtained integrating $\boldsymbol\theta$ out of the likelihood $L(\boldsymbol\theta; \boldsymbol\lambda)$. It is, however, more computationally efficient and equally theoretically founded to maximise the log Laplace approximation 
\begin{align*}
    \ell_{LA}(\boldsymbol\lambda) = \ell(\hat{\boldsymbol\theta}) - \dfrac{1}{2} \hat{\boldsymbol\theta}\ts \textbf{S}_{\boldsymbol\lambda} \hat{\boldsymbol\theta} + \dfrac{1}{2} \log |\textbf{S}_{\boldsymbol\lambda}| - \dfrac{1}{2} \log |-\textbf{H}(\hat{\boldsymbol\theta}) + \textbf{S}_{\boldsymbol\lambda}|,
\end{align*}
where $\hat{\boldsymbol\theta} = \operatorname{arg \ max}_{\boldsymbol\theta} L(\boldsymbol\theta; \boldsymbol\lambda)$ for a given $\boldsymbol\lambda$. At $\boldsymbol\theta^{[a+1]}$, the update for the $k^{th}$ element of $\boldsymbol\lambda^{(rr')}$ for all $(r,r') \in  \mathcal{A}$ is 
\begin{align}\label{eq:SM-EFS}
\lambda_{k}^{(rr')[a+1]} &= \lambda_{k}^{(rr')[a]} \times \frac{\text{tr} \bigg\{  \textbf{S}_{\boldsymbol\lambda^{[a]}}^{-1} \dfrac{\partial \textbf{S}_{\boldsymbol\lambda}}{\lambda_{k}^{(rr')}} \bigg|_{\blambda = \blambda^{[a]}} \bigg\} - \text{tr}\bigg\{ [-\textbf{H}(\hat{\boldsymbol\theta}) + \textbf{S}_{\boldsymbol\lambda^{[a]}}]^{-1} \dfrac{\partial \textbf{S}_{\boldsymbol\lambda}}{\partial \lambda_{k}^{(rr')}} \bigg|_{\blambda = \blambda^{[a]}} \bigg\}}{\hat{\boldsymbol\theta}^\top \bigg( \dfrac{\partial \textbf{S}_{\boldsymbol\lambda}}{\partial \lambda_{k}^{(rr')}} \bigg|_{\blambda = \blambda^{[a]}} \bigg) \hat{\boldsymbol\theta}},
\end{align}
with $k = 1, \dots, K^{(rr')}$. The two steps, (\ref{eq1st1}) and either (\ref{PIRLS1}) or (\ref{eq:SM-EFS}), are iterated until the algorithm satisfies the stopping rule $\frac{\left|\ell(\boldsymbol\theta^{[a+1]}) - \ell(\boldsymbol\theta^{[a]})\right|}{0.1 + \left|\ell(\boldsymbol\theta^{[a+1]})\right|} < 1e-07$, and convergence is assessed by checking that the maximum of the absolute value of the gradient vector is numerically equivalent to 0 and that the observed information matrix is positive definite. In practice, we found the two smoothing methods to yield similar smooth term estimates.   

As with any estimation algorithm, convergence failures may occur. With multi-state models, we mainly found this to be the case when the information provided by the data is insufficient to support the model specified. For instance, when a transition is characterized by a low number of observations, empirical identification of a non-trivial model may not possible. And this tends to be independent of the starting values and of the smoothing method chosen. Such pathological behaviour can often already be spotted in the first few iterations of the optimisation algorithm, with the proposed estimates leading to very large transition intensity values $(> 10^{5})$.

\newpage

\section{Simulation study}\label{suppl:sim-study}

To exemplify the empirical effectiveness of the proposed approach in recovering the true values of key quantities of interest (e.g., transition intensity curves), we carried out two simulation studies. The first one replicates that designed in \cite{machado2021penalised} and uses an IDM set-up. The second study is about a five-state Markov process and serves to illustrate the performance of the proposal in a setting that is more complex than those supported by the methods available in the literature.

\subsection{IDM based simulation}\label{suppl:sim-study-IDM}

We consider a progressive IDM, assuming a different time-dependent shape for each of the three allowed transitions. The time-to-events relating to transition $1 \rightarrow 2$ are simulated from a log-normal distribution with location $1.25$ and scale $1$. This implies that the hazard increases first and then decreases at a later time. For $1 \rightarrow 3$, an exponential distribution with rate $\exp(-2.5)$ is employed. For $2 \rightarrow 3$, we assume a strictly increasing hazard by simulating the time-to-events from a conditional Gompertz distribution with rate $\exp(-2.5)$ and shape $0.1$. For this transition, we have to condition on the event that the individual transitions to state 2 to ensure that the simulated time is larger than the $1 \rightarrow 2$ transition time. As in \cite{machado2021penalised}, we simulate $N = 500$ trajectories (i.e., individuals) $\mathcal{M} = 100$ times. Tests with larger $\mathcal{M}$ confirmed the results reported below, thus we kept $\mathcal{M} = 100$ to retain the comparability with \cite{machado2021penalised}.

More specifically, let $T_{rs} = T_{rs|u}$ represent the time of the transition to state $r'$ conditional on being in state \textit{r} at time $u > 0$. If the state at \textit{u} is 1 then the time of transition to the next state can be obtained by taking $T = \min\{T_{12}, T_{13}\}$. If $T = T_{12}$ then the next state is 2, otherwise the next state is 3. If the state is 2 then the time of the next state is $T_{23}$. Censoring needs to be imposed to render the data intermittently observed; we assume a yearly time-grid spanning over 15 years, i.e. $(t_{i 0}, t_{i 1}, \dots, \min\{t_{i 15}, T_{13}\}) = (0, 1, \dots, \min\{t_{i 15}, T_{13}\})$ for $i = 1, \dots, N$. The reader is referred to \cite{van2016multi} for further details on how to simulate intermittently-observed multi-state survival data. The transition intensities are specified as $q^{(rr')}(t) = \exp \left[ \beta_0^{(rr')} + s_1^{(rr')}(t) \right]$ for $(r,r') \in \{(1,2), (1,3), (2,3)\}$, where $s_1^{(rr')}(t)$ is represented using a cubic regression spline with $J_1^{(rr')} = 10$ and second order penalty. 

\begin{figure}[htb!]
\hspace{0.15cm}
\begin{minipage}{\textwidth}
  \centering
\includegraphics[scale = 0.575, angle = 270]{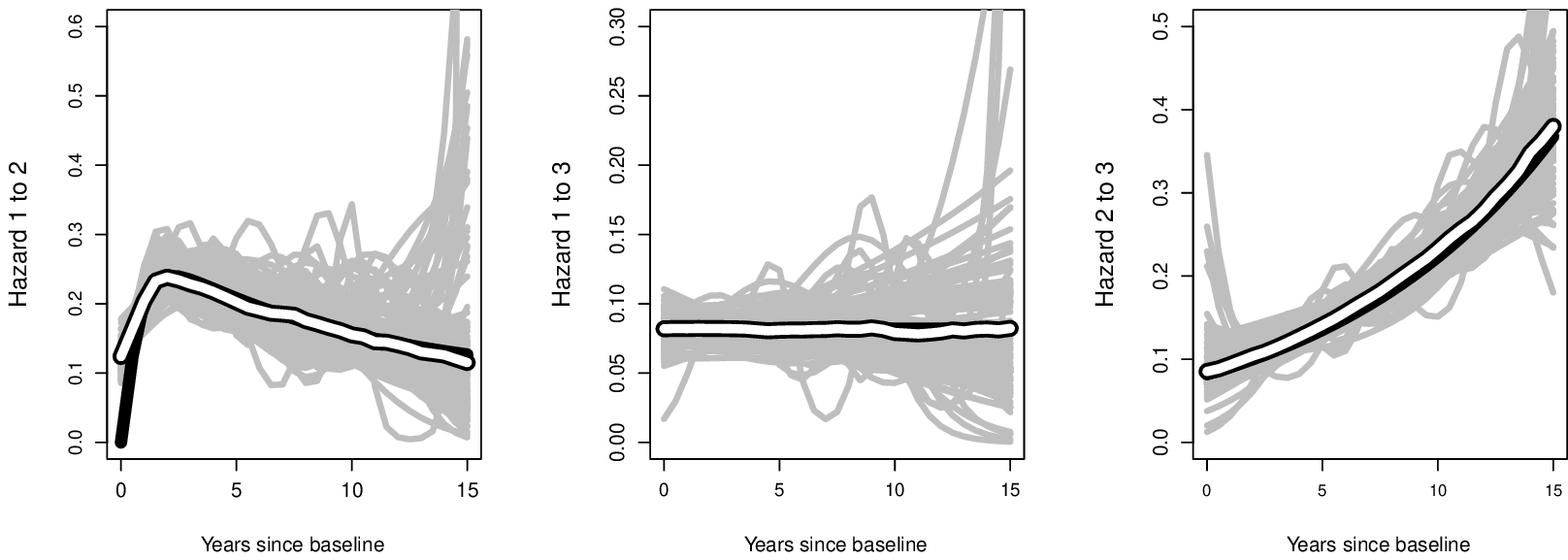}
\end{minipage}
\vfill
\begin{minipage}{\textwidth}
  \centering
\includegraphics[scale = 0.45]{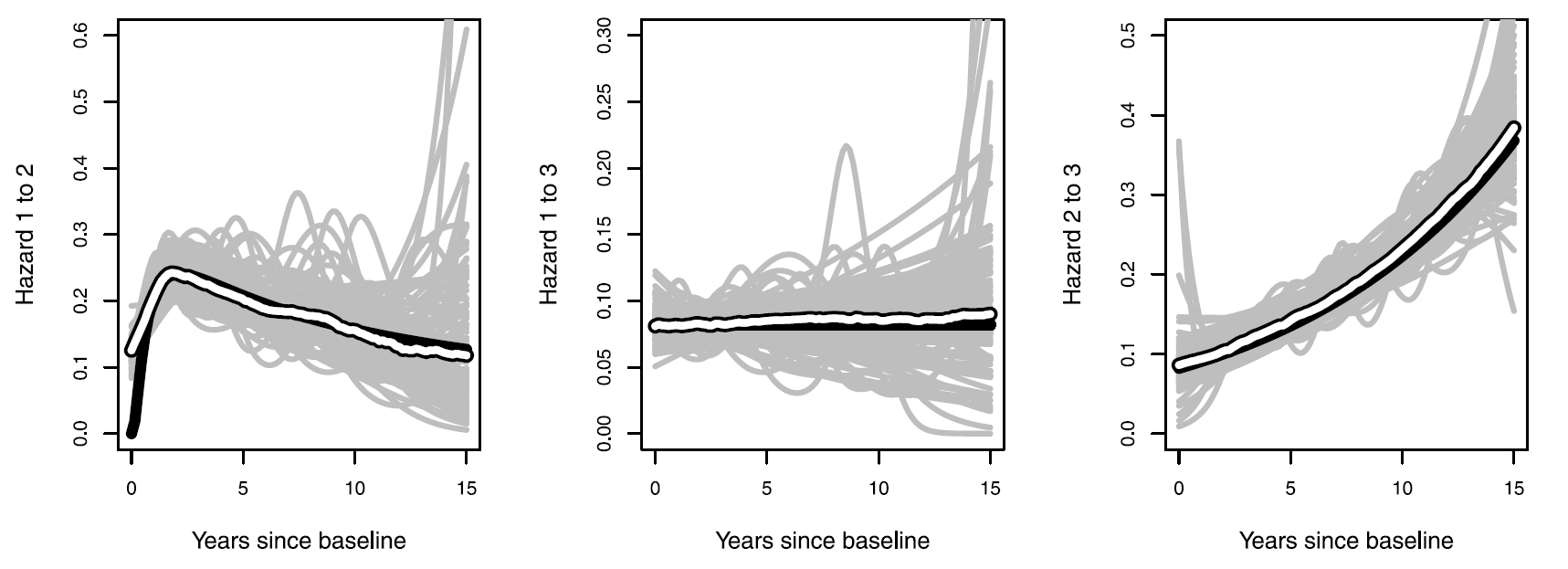}
\end{minipage}
\caption{True (black), estimated (grey, $\mathcal{M}=100$ replicates) and median estimated (white) hazard functions for transitions $1 \rightarrow 2$ (left), $1 \rightarrow 3$ (middle) and $2 \rightarrow 3$ (right) obtained by \texttt{flexmsm} (top row) and \cite{machado2021penalised} (bottom row).}
\label{fig:sim_q_plots}
\end{figure}

In line with \cite{machado2021penalised}, Figure \ref{fig:sim_q_plots} shows the estimated median and true hazards as well as all the $\mathcal{M}$ estimated hazards. Note that the large variation observed towards the end of the study time is due to scarceness of data at later years. Overall, the plots show that the proposed approach is able to recover well the true transition intensity curves for each allowed transition, and that the performance is similar across the two methods. The discrepancy between fitted median and true hazards for transition $1 \rightarrow 2$ is due to definition of interval censoring adopted in the simulation study: the sampling design implies that the living states are observed at intervals of one year; for the first two years after baseline, this design does not work well. 

We also evaluated our approach on the transition probability scale. In particular, Table \ref{tab:tran-prob-sim} reports the true, average and median ten-year estimated transition probabilities, where the average is taken over the $\mathcal{M}$ simulations. The biases are also reported and are defined as $\text{Bias}^{(rr')}(t) = \dfrac{1}{\mathcal{M}} \bigg( \sum\limits_{\nu = 1}^{\mathcal{M}} p^{(\nu, rr')}(0,10) - p^{(rr')}(0,10) \bigg)$, where $p^{(\nu, rr')}(0,10)$ denotes the estimated ten-year probability of transitioning from state $r$ to state $r'$ for the $\nu^{th}$ simulated dataset. Our methodology recovers well the true ten-year transition probabilities and consistently outperforms the approach of \cite{machado2021penalised}.

\begin{table}[htb!]
    \centering
    \begin{tabular}{c | c r | c r }
    \hline
         \multirow{2}{*}{True}    & \multicolumn{2}{c|}{\texttt{flexmsm} (anal.)} & \multicolumn{2}{c}{M. et al. (2021)} \\
                                  & Mean    & Bias     & Mean    & Bias \\
           \hline
         $p^{(11)}(0,10) = 0.065$ & $0.063$ & $-0.002$ & $0.060$ & $0.004$ \\
         $p^{(12)}(0,10) = 0.231$ & $0.232$ & $0.001$  & $0.222$ & $0.009$ \\ 
         $p^{(13)}(0,10) = 0.704$ & $0.705$ & $0.001$  & $0.718$ & $-0.014$ \\
         $p^{(22)}(0,10) = 0.245$ & $0.242$ & $-0.003$ & $0.231$ & $0.014$ \\
         $p^{(23)}(0,10) = 0.755$ & $0.758$ & $0.003$  & $0.769$ & $-0.014$ \\
         \hline
    \end{tabular}
    \caption{Ten-year true and average estimated transition probabilities, and bias for $\mathcal{M} = 100$ replicates.}
    \label{tab:tran-prob-sim}
\end{table}

Finally, we explored the effect that the length of the gap occurring between two successive observations has on estimation performance; it is known that when such gap is large, identifiability issues may arise. To this end, we additionally considered two-, three-, four- and five-yearly time-grids. As expected, the performance deteriorated as the gap increased, with reasonable results (not reported here, but available upon request) still attainable for two- and three-yearly time-grids.

\newpage

\subsubsection{Approximate information matrix}



This section provides some evidence on the convergence performance of the proposed approach when employing an information matrix approximated via first order analytical derivatives. When comparing the results for the analytic Hessian based estimation (M1) and approximate one (M2), we found that the proportions of simulated replicates M1 was better than M2 were:
   \begin{itemize}
        \item $94.5\%$ when analysing the total numbers of iterations for the trust region and smoothing steps discussed in Section \ref{suppl:algorithm};
        \item $81.5\%$ when examining the log-likelihoods of M1 and M2;
        \item $61.5\%$s when comparing the gradients of M1 and M2.
    \end{itemize}

\begin{figure}[htb!]
\hspace{0.15cm}
\begin{minipage}{\textwidth}
  \centering
\includegraphics[scale = 0.5, angle = 270]{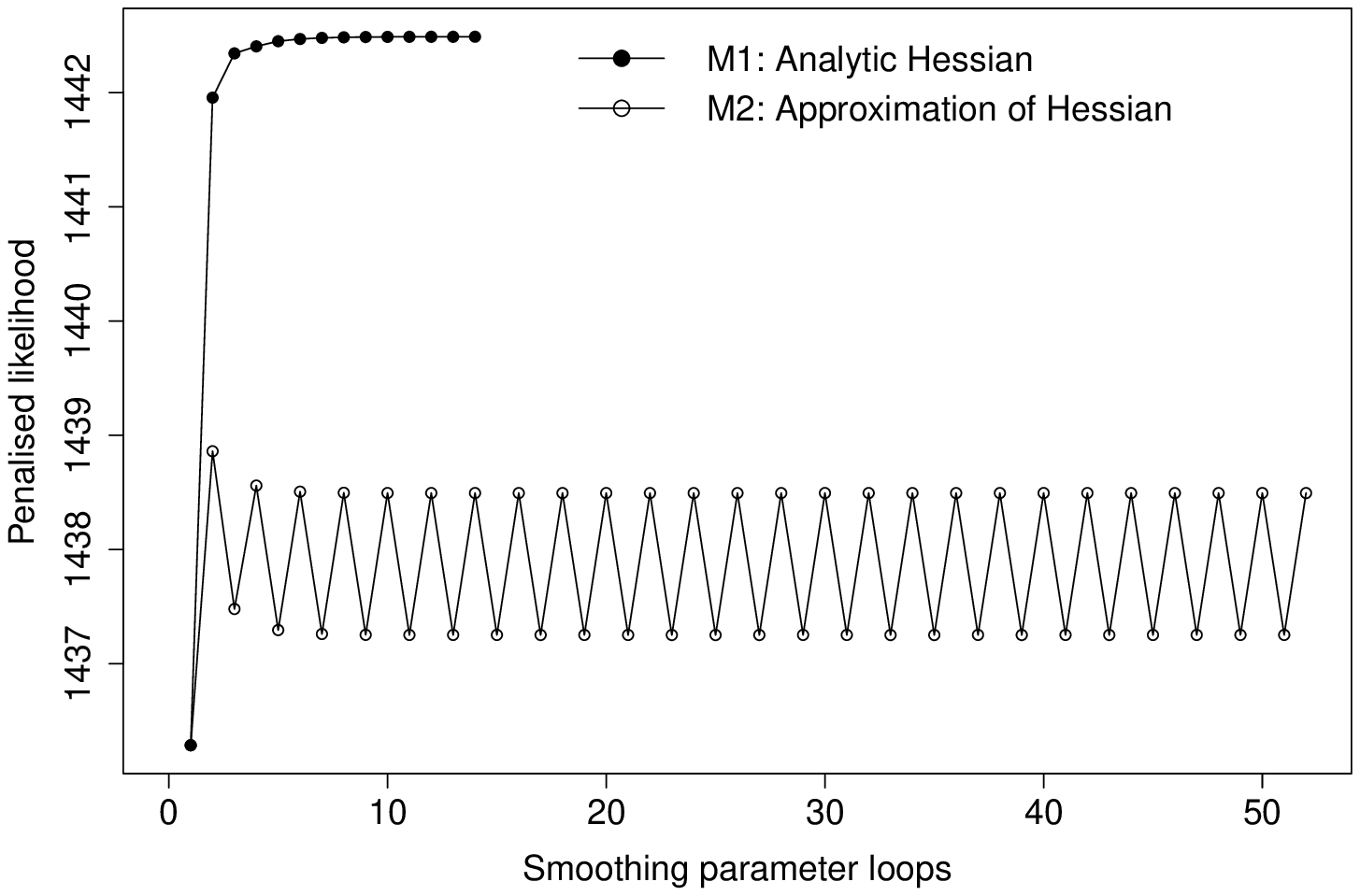}
\end{minipage}
\caption{Penalized log-likelihood at each iteration of the proposed estimation approach, based on M1 and M2, for model (\ref{ours}) in the CAV case study. The run-times on a laptop with Windows 10, Intel 2.20 GHz processor, 16 GB of RAM and eight cores, were 17 minutes for M1 and over 2 hours for M2.} 
\label{fig:liksAnalVsApprox}
\end{figure}

\noindent Note that, in many of the simulated replicates, M2 exhibited a behavior similar to that depicted in Figure \ref{fig:liksAnalVsApprox}, hence highlighting the importance of exploiting in model fitting the information provided by the analytical Hessian matrix of the log-likelihood.  

Finally, we would like to point out that the model specification employed for the simulation set up explored here is simpler than those investigated in the CAV and ELSA studies; as mentioned in the previous section, this was done for comparability with the results of \cite{machado2021penalised}. The difference in performance between M1 and M2 becomes even starker as the complexity of the model specification increases. In fact, in the case studies, it was not possible to estimate the models of interest when basing estimation on the approximate Hessian, because of convergence failures of the type displayed in Figure \ref{fig:liksAnalVsApprox}.

\newpage

\subsection{Five-state process based simulation}\label{suppl:sim-study-five}

We consider a progressive five-state survival process with an absorbing state, and seven transitions whose parameters were chosen to produce intensities similar to those found in the ELSA case study described in Section \ref{sec:ELSA-study}. In particular, we simulate the time-to-events from (conditional) Gompertz distributions with rates and shapes provided for each transition in Table \ref{tab:rates-shapes}. We simulate $N = 500$ trajectories $\mathcal{M} = 100$ times, which are observed for 40 semesters. An intermittently observation scheme is imposed by assuming that individuals are visited every 4 semesters. The time is then brought back to the year scale. This gives counts of pairs of consecutively observed states that are similar to those found in the ELSA case study.

\begin{table}[htb!]
    \centering
    \begin{tabular}{c c c c c c c c}
    \hline
               & $1 \rightarrow 2$ & $1 \rightarrow 5$ & $2 \rightarrow 3$ & $2 \rightarrow 5$ & $3 \rightarrow 4$ & $3 \rightarrow 5$ & $4 \rightarrow 5$   \\
   \hline
         rate  &  $\exp(-2.25)$ & $\exp(-5)$ & $\exp(-2.20)$ & $\exp(-5)$ & $\exp(-2)$ & $\exp(-5)$ & $\exp(-3)$ \\
         shape &  0.06 & 0.02 & 0.05 & 0.09 & 0.01 & 0.02 & 0.04\\
     \hline
    \end{tabular}
    \caption{Rates and shapes for the (conditional) Gompertz distributions generating the transition times in the five-state process based simulation.}
    \label{tab:rates-shapes}
\end{table}

The transition intensities are specified as $q^{(rr')}(t) = \exp \left[ \beta_0^{(rr')} + s_1^{(rr')}(t) \right]$ for\\$(r,r') \in \{(1,2), (1,5), (2,3), (2,5), (3,4), (3,5), (4,5) \}$, where $s_1^{(rr')}(t)$ is represented using a cubic regression spline with $J_1^{(rr')} = 10$ and second order penalty.

\begin{figure}[htb!]
\begin{minipage}{\textwidth}
  \centering
\includegraphics[scale = 0.6, angle = 270]{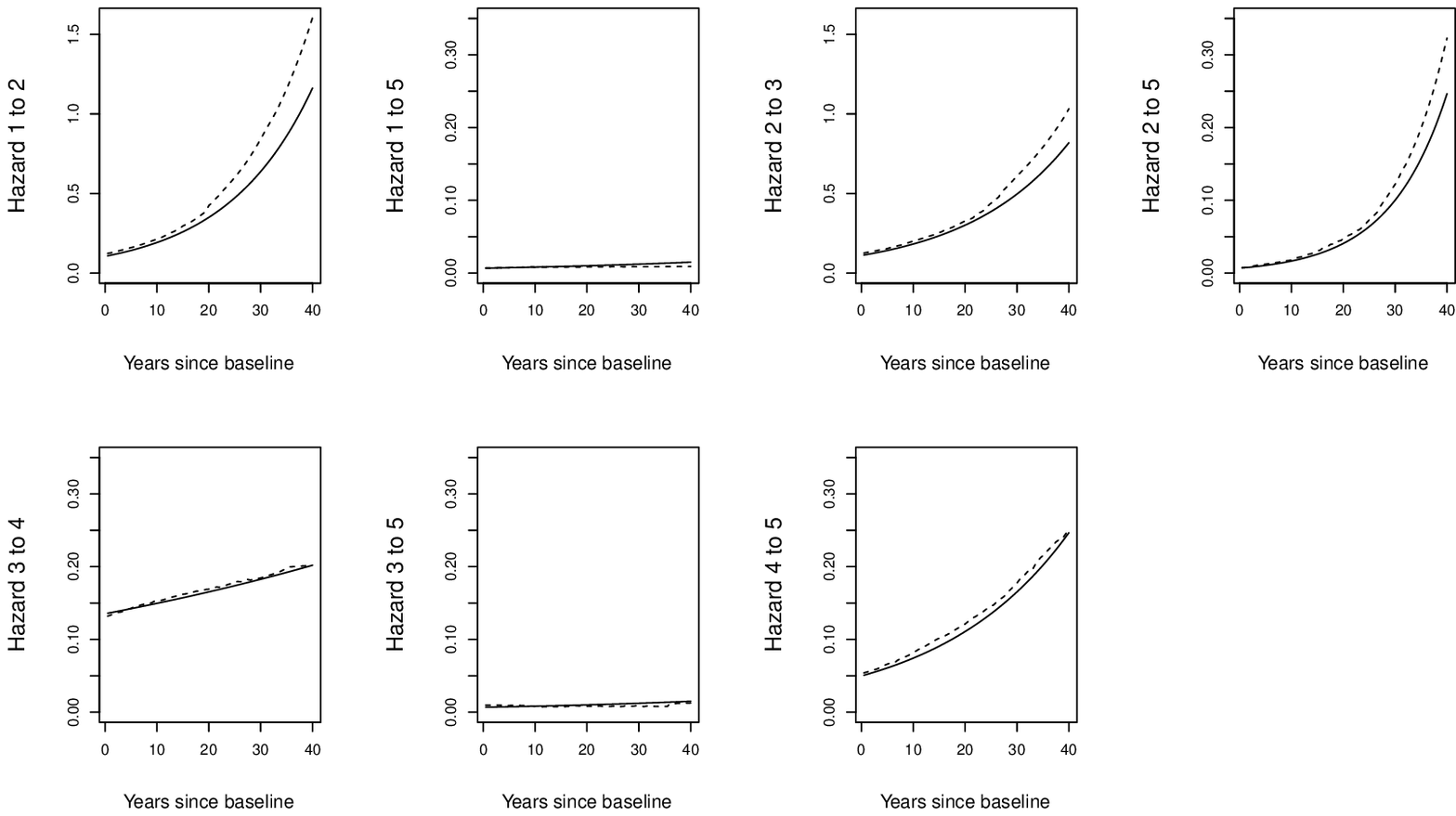}
\end{minipage}
\caption{True (black) and median estimated (dashed) hazard functions for each transition in the simulated five-state process.}
\label{fig:sim_q_5state_plots}
\end{figure}

In Figure \ref{fig:sim_q_5state_plots}, we report the median estimated transition intensities obtained for the $\mathcal{M}$ simulations with our framework, alongside the true curve $q^{(rr')}(t)$, for each of the seven allowed transitions. Overall, the proposed approach recovers adequately the true transition intensity curves.

\begin{table}[htb!]
    \centering
    \begin{tabular}{c | c c }
    \hline
         True & Mean & Bias 
         \\
           \hline
         $p^{(11)}(0,10) = 0.229$ & 0.192 & -0.037 
         \\
         $p^{(12)}(0,10) = 0.318$ &  0.300 & -0.018  
         \\ 
         $p^{(13)}(0,10) = 0.230$ &  0.255 & 0.025 
         \\
         $p^{(14)}(0,10) = 0.121$ &  0.137 & 0.016  
         \\
         $p^{(15)}(0,10) = 0.102$ &  0.116 & 0.014 
         \\
         $p^{(22)}(0,10) = 0.222$ &  0.186 & -0.036  
         \\
         $p^{(23)}(0,10) = 0.330$ &  0.333 & 0.003 
         \\ 
         $p^{(24)}(0,10) = 0.294$ &  0.299 & 0.006 
         \\
         $p^{(25)}(0,10) = 0.154$ &  0.181 & 0.027 
         \\
         $p^{(33)}(0,10) = 0.225$ &  0.222 & -0.003 
         \\
         $p^{(34)}(0,10) = 0.508$ &  0.481 & -0.027 
         \\
         $p^{(35)}(0,10) = 0.267$ &  0.297 & 0.03  
         \\ 
         $p^{(44)}(0,10) = 0.549$ &  0.527 & -0.021  
         \\
         $p^{(45)}(0,10) = 0.451$ &  0.473 & 0.021  
         \\
         \hline
    \end{tabular}
    \caption{Ten-year true, average and median transition probabilities for our framework. The order is that found when reading the transition probability matrix row-wise.}
    \label{tab:tran-prob-sim-5state}
\end{table}

As done for the three-state simulated process, we also evaluate our approach on the transition probabilities scale. In Table \ref{tab:tran-prob-sim-5state}, we report the true and average ten-year estimated transition probabilities, where the average is taken over the $\mathcal{M}$ simulations, and the corresponding biases. The method is able to recover the true ten-year transition probabilities reasonably well, exhibiting consistently small biases. This is reassuring considering the multi-state process adopted here, which is more involved and complex that those commonly explored and used in the literature.

\newpage
\clearpage

\section{List of symbols}\label{suppl:symbols}

\textbf{Covariates and functions or longer terms}
\begin{itemize}
    \item $age_\iota$ covariate in model
    \item $\text{Bias}^{(rr')}(t)$ bias relating to the $r \rightarrow r'$ transition at time $t$ in the simulation study
    \item $\texttt{dage}_{ij}$ covariate in CAV model
    \item $\texttt{pdiag}_{ij}$ covariate in CAV model
    \item $sex_\iota$ covariate in model
    \item $\texttt{sex}_{ij}$ covariate in ELSA model
    \item $\texttt{higherEdu}_{ij}$ covariate in ELSA model
    \item \textit{edf} for effective degrees of freedom
    \item $\text{tr}(\cdot)$ trace function 
    \item $\textbf{1}_{\check{n}}$ vector of 1s of length $\check{n}$.
\end{itemize}

\vspace{1cm}

\textbf{Latin letters}
\begin{itemize}
    \item $a$ estimation algorithm iteration index.
    \item $\textbf{A}$ matrix of eigenvectors.
    \item $\mathcal{A}$  set of allowed transitions
    \item $\textbf{b}_{k}^{(rr')}(\tilde{\textbf{x}}_{k \iota})$ bases function vector for the $k^{th}$ term in the $(r,r')$ transition intensity.
    \item $c$ indexing for likelihood contributions (censored state contribution and for exactly observed absorbing state).
    \item $C$  total number of states.
    \item $d_\upsilon$ difference of knots in the construction of the cubic regression spline.
    \item $\textbf{D}_k^{(rr')}$ penalty matrix for the $k^{th}$ term in the $(r,r')$ transition intensity.
    \item $\textbf{e}$ vector in the Taylor approximation.
    \item $\mathbf{E}$ matrix found in the closed-form expressions of the first and second derivatives of the transition probability matrix.
    \item $\mathbb{E}$ expectation function.
    \item $f_{\boldsymbol{\theta}}$ prior on the model parameter $\boldsymbol{\theta}$.
    \item $G_{lm}^{(w)}$ the $(l,m)$ element of $\textbf{G}^{(w)}$.
    \item $G_{lm}^{(w w')}$ the $(l,m)$ element of $\textbf{G}^{(w w')}$.
    \item $\textbf{g}(\boldsymbol{\theta})$ gradient vector.
    \item $\textbf{G}^{(w)}$ matrix needed for the closed form expression of $\partial^2 \textbf{P}$ (transformation of first derivative of Q matrix).
    \item $\textbf{G}^{(w w')}$ matrix needed for the closed form expression of $\partial^2 \textbf{P}$ (transformation of second derivative of Q matrix).
    \item $h$ infinitesimal time in the limit-based definition of the transition intensity.
    \item $\textbf{H}(\boldsymbol{\theta})$ hessian matrix. 
    \item $\textbf{H}_p(\boldsymbol{\theta})$ penalized hessian matrix.
    \item $i$ indexing for the statistical units when defining the likelihood. Here $i = 1, \dots, N$.
    \item $j$ indexing for the observations of a specific statistical unit.
    \item $J_k^{(rr')}$ number of basis functions for the $k^{th}$ term in $(r,r')$ transition intensity.
    \item $k$ indexing for overall covariate/parameter vector, with $k = 1, \dots, K^{(rr')}$.
    \item $K^{(rr')}$ total number of terms in additive predictor $\eta_i^{(rr')}(t_i, \x_i;\boldsymbol{\beta}^{(rr')})$, excluding the intercept.
    \item $l$ indexing for the $(l,m)$ element of the matrices needed for the closed form expression of $\partial^2 \textbf{P}$.
    \item $\ell_{LA}$ log Laplace approximation of $L(\boldsymbol{\lambda})$.
    \item $\ell(\boldsymbol\theta)$ model log-likelihood.
    \item $\ell_p(\boldsymbol{\theta})$ penalized log-likelihood.
    \item $\breve{\ell}_p(\boldsymbol\theta)$ second order approximation of the model log-likelihood.
    \item $L_{ij}(\boldsymbol{\theta})$ likelihood contribution for $j^{th}$ observation of $i^{th}$ individual.
    \item $L(\boldsymbol{\theta}; \boldsymbol{\lambda})$ joint log density (used to explain efs smoothing approach).
    \item $L(\boldsymbol{\lambda})$ joint log density when integrating out $\boldsymbol{\theta}$ (used to explain efs smoothing approach).
    \item $m$ indexing for the $(l,m)$ element of the matrices needed for the closed form expression of $\partial^2 \textbf{P}$.
    \item $\mathcal{M}$ number of simulations in the simulation study.
    \item $\textbf{M}$ matrix appearing in the update of the smoothing parameter.
    \item $N$ total number of statistical units. 
    \item $\check{n}$ total number of observations in the dataset.
    \item $n_i$ number of observations for the $i^{th}$ statistical unit with $i = 1, \dots, N$.
    \item $n_{sim}$ number of simulations used to obtain confidence intervals.
    \item $\textbf{O}$ quantity appearing in the smoothing parameter update and \textit{edf} definition.
    \item $p^{(rr')}(t,t')$ transition probabilities referring to time interval $(t,t')$.
    \item $p^{(\nu, rr')}(t,t')$ the $\nu^{th}$ simulated transition probability referring to time interval $(t,t')$, with $\nu = 1, \dots, \mathcal{M}$.
    \item $\textbf{P}(t,t')$ transition probability matrix referring to time interval $(t,t')$.
    \item $\hat{\textbf{P}}(t,t')$ estimated transition probability matrix referring to time interval $(t,t')$.
    \item $q^{(rr')}(t)$ transition intensity at time $t$.
    \item $q^{(n_{sim}, rr')}$ the $n_{sim}^{th}$ simulated transition intensity (for confidence interval construction).
    \item $\textbf{Q}(t)$ transition intensity matrix at time $t$.
    \item $\hat{\textbf{Q}}(t)$ estimated transition intensity matrix at time $t$.
    \item $\textbf{Q}_j(\boldsymbol{\theta})$ transition intensity matrix at the $j^{th}$ observation of a generic individual. 
    \item $r$  starting state. 
    \item $r'$ arrival state.
    \item $\mathbb{R}$ real numbers set.
    \item $s_k^{(rr')}(\tilde{\textbf{x}}_{k \iota})$ $k^{th}$ smooth for the $(r,r')$ transition intensity.
    \item $\mathcal{S}$  state space of process.
    \item $\textbf{S}^{(rr')}_{\boldsymbol{\lambda}^{(rr')}}$ penalty term for the $(r,r')$ transition intensity.
    \item $\textbf{S}_{\boldsymbol{\lambda}}$ overall penalty term.
    \item $t$ and $t'$ generic time.
    \item $t_{ij}$ with $i = 1, \dots, N$ and $j = 1, \dots, n_i$ is the $j^{th}$ observed time for the $i^{th}$ statistical unit.
    \item $t_j$ used as shorthand of $t_{ij}$ for the generic statistical unit (i.e. when dropping $i$ for simplicity).
    \item $\delta t$ time interval in the definition of the closed form expression of $\textbf{P}$
    \item $T_{rs}$ time of the $r \rightarrow r'$ transition
    \item $T_{rs\mid u}$ time of the $r \rightarrow r'$ transition conditional on being in state $r$ at time $u$
    \item $u$ integration variable when integrating transition intensity.
    \item $u_\upsilon$ knot for the example in the (cubic regression) smooth of time.
    \item $\check{\textbf{U}}_{w w'}$ one of the matrices of the closed form expression of $\dfrac{\partial^2 }{\partial \theta_w \partial \theta_{w'}} \textbf{P}$.
    \item $\dot{\textbf{U}}_{w}$ one of the matrices of the closed form expression of $\dfrac{\partial}{\partial \theta_w} \textbf{P}$.
    \item $\dot{\textbf{U}}_{w w'}$ one of the matrices of the closed form expression of $\dfrac{\partial^2 }{\partial \theta_w \partial \theta_{w'}} \textbf{P}$.
    \item $\textbf{V}_{\boldsymbol{\theta}}$ estimated negative inverse penalized Hessian.
    \item $w$ and $w'$ indexing for gradient vector and Hessian, with $w, w' = 1, \dots, W$.
    \item W total number of parameters
    \item $\textbf{x}_i$ covariate vector (without time).
    \item $\tilde{\textbf{x}}_\iota$ overall covariate vector (with time).
    \item $\tilde{\textbf{x}}_{k \iota}$ is the $k^{th}$ sub-vector of the overall covariate vector $\textbf{z}_i$.
    \item $\tilde{\textbf{X}}_{k}^{(rr')}$ the design matrix corresponding to the $k^{th}$ term in the $(r,r')$ transition intensity.
    \item $\tilde{\textbf{X}}^{(rr')}$ overall design matrix for the $(r,r')$ transition intensity.
    \item $y$ indexing of the eigenvalues.
    \item $Y$ number of eigenvalues.
    \item $z_{ij}$ with $i = 1, \dots, N$ and $j = 1, \dots, n_i$ is the $j^{th}$ state occupied by the $i^{th}$ statistical unit. 
    \item $Z(t)$  multi-state process.
\end{itemize}

\vspace{1cm}

\textbf{Greek letters}
\begin{itemize}
    \item $\alpha$ confidence level.
    \item $\beta_0^{(rr')}$ intercept parameter for $(r,r')$ transition intensity.
    \item $\boldsymbol{\beta}_k^{(rr')}$ parameter vector for the $k^{th}$ term in the $(r,r')$ transition intensity. Its length is $J_k^{(rr')}$.
    \item $\boldsymbol{\beta}^{(rr')}$ parameter vector for $(r,r')$ transition intensity. Its length is $\sum_{k = 1}^{K^{(rr')}} J_k^{(rr')}$. 
    \item $\hat{\boldsymbol{\beta}}^{(rr')}$ estimated parameter vector of $\boldsymbol{\beta}^{(rr')}$.
    \item $\boldsymbol{\beta}^{(n_{sim}, rr')}$ the $n_{sim}^{th}$ simulated parameter vector for the $(r,r')$ transition intensity.
    \item $\gamma_y$ the $y^{th}$ eigenvalue, with $y = 1, \dots, Y$.
    \item $\boldsymbol{\Gamma}$ matrix of eigenvalues.
    \item $\delta t$ time interval in the definition of the closed form expression of the transition probability matrix (and its derivatives).
    \item $\Delta^{[a]}$ radius of the trust region at the $a^{th}$ iteration.
    \item $\boldsymbol{\epsilon}$ quantity appearing in the smoothing parameter update.
    \item $\zeta$ indexing for the series representing the exponential.
    \item $\eta_\iota^{(rr')}(t_\iota, \x_\iota;\boldsymbol{\beta}^{(rr')})$ additive predictor.
    \item $\boldsymbol{\eta}^{(rr')}$ overall additive predictor for the $(r,r')$ transition intensity.
    \item $\boldsymbol{\theta}$ overall parameter vector.
    \item $\hat{\boldsymbol{\theta}}$ estimated overall parameter vector.
    \item $\boldsymbol{\theta}^{[a]}$ overall parameter vector at the $a^{th}$ iteration of the estimation algorithm.
    \item $\iota$ indexing of the observations when defining the additive predictor. Here $i = 1, \dots, \check{n}$.
    \item $\kappa$ indexing for the summations appearing in the proof of the $\partial^2 \textbf{P}$ expression.
    \item $\lambda_k^{(rr')}$ smoothing parameter for the $k^{th}$ term in the $(r,r')$ transition intensity.
    \item $\boldsymbol\lambda^{(rr')}$ smoothing parameter vector in the $(r,r')$ transition intensity. It's length is $K^{(rr')}$.
    \item $\boldsymbol{\lambda}$ overall smoothing parameter vector.
    \item $\boldsymbol{\mu}_{\textbf{M}}$ and $\hat{\boldsymbol{\mu}}_{\textbf{M}}$ quantity appearing in the smoothing parameter update.
    \item $\nu$ indexing for simulated probabilities to compute the bias in the simulation study.
    \item $\rho$ indexing for the summations appearing in the proof of the $\partial^2 \textbf{P}$ expression.
\end{itemize}

\end{document}